\documentclass[11pt]{article}

\usepackage{amsthm,amsmath,amssymb,amsfonts,bbm,bm}
\usepackage{slashed}

\usepackage{tocloft} 

\usepackage[nosort]{cite} 
\usepackage{graphicx,color}
\usepackage[colorlinks=true, linkcolor=blue, citecolor=blue, linktoc=all]{hyperref}

\usepackage[usenames,dvipsnames]{xcolor}

\usepackage{tikz-cd}
\usepackage{pict2e}
\usepackage{physics}
\usepackage{comment} 
\usepackage[makeroom]{cancel}

\newcommand{\beq}{\begin{eqnarray}}
\newcommand{\eeq}{\end{eqnarray}}
\newcommand{\beqn}{\begin{eqnarray}}
\newcommand{\eeqn}{\end{eqnarray}}
\newcommand{\bea}{\begin{eqnarray}}
\newcommand{\eea}{\end{eqnarray}}
\newcommand{\be}{\begin{equation}}
\newcommand{\ee}{\end{equation}}

\newcommand{\ack}[1]{[{\bf Pfft!: {#1}}]}

\newcommand{\un}[1]{\underline{#1}}

\usepackage[normalem]{ulem}

\newcommand{\RR}{\mathbb{R}}

\theoremstyle{definition}

\newtheorem{example}{Example}[section]
\newtheorem{definition}{Definition}[section]

\newcommand{\myfig}[3]{
	\begin{figure}[h]
	\centering
	\includegraphics[width=#2cm]{#1}\caption{{\scriptsize #3}}\label{fig:#1}
	\end{figure}
	}

\newcommand{\uiuc}[1]{
	\centerline{
		\begin{minipage}[c]{0.7\textwidth}
			\begin{center}
			${}^{#1}$ Illinois Center for Advanced Studies of the Universe \& Department of Physics,\\ 
			University of Illinois, 1110 West Green St., Urbana IL 61801, U.S.A.
			\end{center}
		\end{minipage}
		}
	}

\usepackage{mathrsfs,mathtools}
\renewcommand\mathbb[1]{\mathbbm{#1}}

\newcommand{\td}{{\rm d}}

\newcommand{\hatd}{\hat{d}}

\newcommand{\mX}{\un{\mathfrak{X}}}

\newcommand{\mh}{\mathfrak{h}}

\makeatletter
\DeclareRobustCommand{\loplus}{\mathbin{\mathpalette\dog@lsemi{+}}}
\DeclareRobustCommand{\lotimes}{\mathbin{\mathpalette\dog@lsemi{\times}}}
\DeclareRobustCommand{\roplus}{\mathbin{\mathpalette\dog@rsemi{+}}}
\DeclareRobustCommand{\rotimes}{\mathbin{\mathpalette\dog@rsemi{\times}}}

\newcommand{\dog@rsemi}[2]{\dog@semi{#1}{#2}{-90,90}}
\newcommand{\dog@lsemi}[2]{\dog@semi{#1}{#2}{270,90}}
\newcommand{\dog@semi}[3]{%
  \begingroup
  \sbox\z@{$\m@th#1#2$}%
  \setlength{\unitlength}{\dimexpr\ht\z@+\dp\z@\relax}%
  \makebox[\wd\z@]{\raisebox{-\dp\z@}{%
    \begin{picture}(1,1)
    \linethickness{\variable@rule{#1}}
    \roundcap
    \put(0.5,0.5){\makebox(0,0){\raisebox{\dp\z@}{$\m@th#1#2$}}}
    \put(0.5,0.5){\arc[#3]{0.5}}
    \end{picture}%
  }}%
  \endgroup
}
\newcommand{\variable@rule}[1]{%
  \fontdimen8  
  \ifx#1\displaystyle\textfont3\else
    \ifx#1\textstyle\textfont3\else
      \ifx#1\scriptstyle\scriptfont3\else
        \scriptscriptfont3\relax
  \fi\fi\fi
}
\DeclareRobustCommand{\loplus}{\mathbin{\mathpalette\dog@lsemi{+}}}

\newcommand{\thistitle}{Crossed Products, Extended Phase Spaces\\ and the Resolution of Entanglement Singularities}

\begin{document}

\title{\thistitle}
\author{
	Marc S. Klinger 
	and 
	Robert G. Leigh 
	\\
	\\
	{\small \emph{\uiuc{}}}
	\\
	}
\date{}
\maketitle
\vspace{-0.5cm}
\begin{abstract}
\vspace{0.3cm}
We identify a direct correspondence between the crossed product construction which plays a crucial role in the theory of Type III von Neumann algebras, and the extended phase space construction which restores the integrability of non-zero charges generated by gauge symmetries in the presence of spatial substructures. This correspondence provides a blue-print for {\it resolving} singularities which are encountered in the computation of entanglement entropy for subregions in quantum field theories. The extended phase space encodes quantities that would be regarded as `pure gauge' from the perspective of the full theory, but are nevertheless necessary for gluing together, in a path integral sense, physics in different subregions. These quantities are required in order to maintain gauge covariance under such gluings. The crossed product provides a consistent method for incorporating these necessary degrees of freedom into the operator algebra associated with a given subregion. In this way, the extended phase space completes the subregion algebra and subsequently allows for the assignment of a meaningful, finite entropy to states therein. 
\end{abstract}

\newpage

\begingroup
\hypersetup{linkcolor=black}
\tableofcontents
\endgroup
\noindent\rule{\textwidth}{0.6pt}

\setcounter{footnote}{0}
\renewcommand{\thefootnote}{\arabic{footnote}}
\newpage

\newcommand{\curr}[1]{\mathbb{J}_{#1}}
\newcommand{\constr}[1]{\mathbb{M}_{#1}}
\newcommand{\chgdens}[1]{\mathbb{Q}_{#1}}
\newcommand{\spac}[1]{S_{#1}}
\newcommand{\hyper}[1]{\Sigma_{#1}}
\newcommand{\chg}[1]{\mathbb{H}_{#1}}
\newcommand{\ThomSig}[1]{\hat{\Sigma}_{#1}}
\newcommand{\ThomS}[1]{\hat{S}_{#1}}
\newcommand{\discuss}[1]{{\color{red} #1}}

\section{Introduction}

Of great current interest is the quest to understand quantum field theories (and eventually quantum gravity) on subregions. One approach to this problem is via algebraic quantum field theory \cite{Haag:1963dh}, and there has been much recent discussion of the corresponding operator algebras, particularly in the context of AdS-holography \cite{Leutheusser:2021frk,Leutheusser:2022bgi}. With that being said, this perspective is clearly of more general interest in quantum field theory and quantum gravity. Another avenue of attack involves the so-called 
universal corner symmetry group (UCS) 
\cite{Ciambelli:2021vnn,Ciambelli:2021nmv,Ciambelli:2022vot,Freidel:2021dxw,Ciambelli:2022cfr}
and the closely related notion of extended phase space \cite{Ciambelli:2021vnn,Klinger:2023qna} in gauge theories and diffeomorphism-invariant theories.
Both of these approaches present general strategies for moving beyond the shortcomings of naive quantum field theory. 
It has perhaps not been clear, however, whether these two approaches are related to each other. The primary focus of this note is to motivate the relationship between these two approaches, and to use the intersection between their ideas to formulate a proposal for resolving the entanglement singularities which are encountered in subregion quantum field theories. 

Beginning on the operator algebra side, it is a well known dictum that the operators associated with a subregion of spacetime form a type III von Neumann algebra, $\mathcal{M}_{\mathcal{R}}$ \cite{araki:1964, Longo:1982zz, Fredenhagen:1984dc}. This means that $\mathcal{M}_{\mathcal{R}}$ lacks a semi-finite trace, and does not admit conventional density operators, resulting in the inability to compute finite entanglement observables associated to the subregion. On the other hand, operator algebra theory presents an elegant solution to the lack of conventional tools in a type III von Neumann algebra in the form of Tomita-Takesaki theory \cite{Takesaki:1970aki, Combes1971}. This theory presents an alternative to traces and density operators in the form of faithful, semi-finite weights and the modular automorphism group of $\mathcal{M}_{\mathcal{R}}$. Together, these tools allow for the computation of finite expectation values for well-behaved operators, and for the computation of finite relative entropies between faithful semi-finite states.\footnote{For an introduction to modular theory aimed at physics audiences see, for example, \cite{Borchers:2000pv,Witten_2018}.}

An enlarged perspective on Tomita-Takesaki theory can be reached by considering the crossed product construction \cite{nakamura1958some,turumaru1958crossed}. Given a von Neumann algebra $\mathcal{M}$ acted upon by a group $G$ by means of a homomorphism\footnote{Here $\text{Aut}(\mathcal{M})$ is the set of automorphisms of $\mathcal{M}$.} $\alpha: G \rightarrow \text{Aut}(\mathcal{M})$, the crossed product of $\mathcal{M}$ by $G$ is a von Neumann algebra that, roughly speaking, combines the algebras $\mathcal{M}$ and $G$ along with a faithful action of $G$ on $\mathcal{M}$. Recall that a von Neumann algebra can always be realized as a subalgebra of the bounded operators acting on some Hilbert space by means of a faithful representation $\rho: \mathcal{M} \rightarrow \mathcal{B}(\mathcal{H})$. In light of this fact, an automorphism of $\mathcal{M}$ can be realized as a family of unitary conjugations: $\rho(x) \mapsto U(g) \rho(x) U(g)^*$ for each $x \in \mathcal{M}$, where here $U(g) \in \mathcal{U}(\mathcal{H})$ is a unitary element of $\mathcal{B}(\mathcal{H})$ for each $g \in G$. The automorphism $\alpha$ is called \emph{inner} if $U(g) \in \text{im}(\rho)$ for all $g \in G$, i.e., if the unitary elements carrying out the action of $G$ on $\mathcal{M}$ can be regarded as contained in $\mathcal{M}$ itself. Otherwise, the automorphism $\alpha$ is called \emph{outer}. The automorphism $\alpha$ applied to the crossed product algebra is inner by construction; indeed, one interpretation of the crossed product is as an extension of the von Neumann algebra $\mathcal{M}$ for which the automorphism $\alpha$ is rendered an inner automorphism. 

The fact that the automorphism $\alpha$ is always inner relative to its associated crossed product becomes very significant in the study of type III von Neumann algebras. This is because Takesaki made the crucial observation that any von Neumann algebra for which the modular automorphism group can be realized as an inner automorphism must be semi-finite, and is therefore at least type II. Thus, the crossed product of any von Neumann algebra with its own modular automorphism group is automatically semi-finite. In light of this observation, Takesaki proposed the crossed product of a type III von Neumann algebra with its modular automorphism group as a unique type II von Neumann algebra representing the original type III algebra \cite{takesaki:1973, takesaki1973crossed, Connes1973}. The introduction of the crossed product provides an important extension to the usual Tomita-Takesaki theory. While Tomita-Takesaki theory allows for the computation of finite relative quantities like the relative entropy, the crossed product algebra introduces the possibility of assigning finite entanglement entropies to states on a type III algebra. This assignment follows from the dual weight theorem, which provides a mapping from faithful weights on the original type III algebra into faithful weights on its associated crossed product \cite{Haagerup1978I, Haagerup1978II}. Because the crossed product algebra is semi-finite, this allows for the original type III weights to be associated with density operators in the crossed product algebra. In physical contexts, this suggests that the crossed product and the dual weight theorem allow for a natural entanglement entropy to be assigned to the states of a subregion algebra.

Modular theory and the crossed product therefore provide the necessary tools in order to study subregion operator algebras. However, in many ways they give only a limited physical intuition behind their usefulness. For example, it is perhaps not clear what role regulators play in these fully algebraic discussions. This is important because, traditionally, quantum field theories are only defined in the presence of  a suitable UV regulator. Worse yet, new divergences appear in the computation of entanglement observables; a sign that yet more elaborate regulators are needed to curtail these more exotic singularities. In quantum field theories, one often interprets a large momentum cutoff as a stand-in for some better description, such as an ultraviolet completion. Indeed, in many physical situations, singularities signal the presence of a symmetry enhancement together with the inappropriate elimination of the associated degrees of freedom that have become `light' or physical. It seems natural to propose that there is a similar notion in the context of entanglement singularities. In this note we will argue that this is one way to interpret the crossed product, which completes the naive subregion algebra by including new degrees of freedom associated with the modular automorphism group. Building on that point, we will demonstrate that the relevant physical mechanism underlying the crossed product is generically affiliated with the universal corner symmetry group. In other words, taking into account the physicality of the UCS at corners such as entanglement cuts can be interpreted as providing a {\it resolution} (in contrast to an imposed hard cutoff) of the singularities associated with entanglement observables. As indicated, this provides a physical interpretation behind the usefulness of modular theory and the crossed product.  

Let us discuss the entanglement singularity problem again but now from a more geometric perspective. It has long been appreciated that the quantum physics of subregions in gauge theories and diffeomorphism-invariant theories is complicated by the fact that typical observables are not local. For example, consider a lattice gauge theory in which there are degrees of freedom which reside at each lattice site, but also degrees of freedom that are shared across the links \emph{between} lattice sites \cite{Donnelly:2011hn}. For this reason, the Hilbert space of such a gauge theory resists a naive factorization when the lattice is split into disjoint subregions. In a continuum QFT we observe an analogous phenomenon. The Hilbert space defined on a Cauchy surface does not factorize on subregions due to degrees of freedom which exist within the interface between a subregion and its causal complement. In \cite{Donnelly:2016auv} it was recognized that the physical role of these degrees of freedom is to facilitate the consistent quantum mechanical gluing of subregions. In light of this fact, the authors introduced a fusion product which explicitly takes these degrees of freedom into account, and in turn realized a tensor factorizable Hilbert space. The authors referred to this Hilbert space, which includes the entangling degrees of freedom in addition to the naive subregion Hilbert spaces, as the \emph{extended} Hilbert space. They also outlined how one might compute entanglement observables by passing states in the naive Hilbert space to this extended Hilbert space. There are by now numerous examples in which this has been worked out explicitly (for example in the topological context, see \cite{Fliss:2017wop,Fliss:2020cos}).

 Generally speaking, the significance of  corner symmetries was understood via a careful application of Noether's second theorem to classical field theories on subregions, $\mathcal{R}$, with boundaries, $\partial \mathcal{R} = S$ \cite{noether1971invariant}. By a corner, we mean any codimension-2 surface; here, the corner of interest is the boundary of a subregion which might be an entanglement cut. 
Naively, Noether's second theorem has been interpreted as stating that the charges associated to any gauge symmetry are identically zero. On the contrary, however, it merely dictates that, on-shell, the current density associated with a gauge symmetry is a total derivative. Applying Stokes' theorem, one therefore finds that the charges associated with such a current may be non-zero, and are non-trivially impacted by the geometric embedding of $\mathcal{R}$ and its associated corner $S$ \cite{Ciambelli:2021vnn,Ciambelli:2021nmv,Ciambelli:2022vot,Freidel:2021dxw,Ciambelli:2022cfr,Donnelly:2016auv,Freidel:2023bnj,Balachandran:1994up,Carlip:1994gy,Carlip:1996yb,Balachandran:1995qa,REGGE1974286,Donnelly:2020xgu,Geiller:2017xad,Freidel:2020xyx,Freidel:2020svx,Freidel:2020ayo,Chandrasekaran:2021hxc,Donnelly:2022kfs}. In particular, it was realized in \cite{Ciambelli:2021vnn,Ciambelli:2022cfr,Freidel:2021cjp,Speranza:2017gxd} that there is a closed subalgebra of diffeomorphisms which could generate non-zero charges in the presence of subregions -- predictably, this algebra is the maximal subalgebra of $\text{Diff}(M)$ compatible with the embedding of the corner $S$. This is the universal corner symmetry algebra. One can view this algebra as generating a tubular neighborhood of the corner, as we will recall later. This interpretation was first explained in Ref. \cite{Klinger:2023qna} where we also generalized this algebra to include internal gauge symmetries as well. The origin of the extended Hilbert space has subsequently been traced back to the extension of the symplectic structure preceding quantization when gauge symmetries rendered physical by the presence of spacetime substructures produce non-integrable charges.\footnote{That is, they are no longer faithfully represented by Hamiltonian functions.} In the absence of such an extension the operators associated with these symmetries cannot be accounted for during quantization. From the point of view of operator algebra theory, this has been described as an `incompleteness' \cite{Casini:2021zgr,Witten:2023qsv} of the subregion algebras referred to above, in that they apparently do not include any such boundary configurations or a description of the relevant symmetries.\footnote{There is a close relation here to higher symmetries, a subject that we will return to in a separate publication.} Recent literature often has applied operator algebra theory to contexts in which inclusions of subalgebras are relevant (related to various monotonicity properties). Here we emphasize that just as  Hilbert space  does not generally factorize on subregions, neither do corresponding operator algebras. If one wants to describe at the algebraic level how to glue subregions together, then one must extend the algebras, as we will describe, in a way that involves the crossed product. 

In \cite{Ciambelli:2021nmv, Klinger:2023qna} we have constructed an extended phase space for which these new physical symmetries are realized faithfully. As we have addressed in that work and has been discussed elsewhere \cite{Donnelly:2011hn,Speranza:2017gxd,wald1993black,iyer1994some,Fliss:2017wop,Banados:1992wn,Strominger:1997eq,Donnelly:2014fua,Donnelly:2014gva,Donnelly:2015hxa,Das:2015oha,Wen:2016snr,Carlip:2017xne,Chen:2020nyh,Geiller:2019bti,faulkner2016modular,faulkner2016shape,
balasubramanian2017multi,balasubramanian2018entanglement,Chandrasekaran:2020wwn,Fliss:2020cos}, the degrees of freedom which are added into the extended phase space have a very clear interpretation as being precisely the same degrees of freedom that govern the gluing of subregions, and therefore carry the entanglement between subregions. For example, in \cite{Fliss:2020cos} it was recognized that moving from the naive to the extended Hilbert space in the context of Chern-Simons gauge theories with interfaces corresponds to admitting new gauge \emph{non}-invariant states. The gluing degrees of freedom subsequently describe how gauge parameters are passed through the interface consistently from one subregion to the next to ensure that bulk gauge invariance is restored. These degrees of freedom have an explicit representation as gauge charges which act like global symmetry generators on the interface. In this respect it has long been hypothesized that extending the phase space must in some way play a similar role to the modular theory, rendering entanglement observables accessible to the theory. This is the point that we wish to formalize.

In this note we propose a correspondence between the crossed product and the extended phase space. Our point of view is that these two procedures are in a sense classical and quantum duals of each other. In the former, one begins with a von Neumann algebra $\mathcal{M}$ acted upon by a symmetry group $G$ and seeks to incorporate the generators of the symmetry into the algebra by means of an automorphism. In the latter, one begins with a (pre-) symplectic manifold $(X,\Omega)$ with a presymplectic action by a group $G$ and seeks to represent the generators of the symmetry as functions on the symplectic geometry by means of an equivariant moment map. In both constructions, one distinguishes between cases in which the symmetry generators can be incorporated with or without the need to augment the original structure. In the case of $\mathcal{M}$, this corresponds to whether the automorphism is inner or outer. In the case of $(X,\Omega)$, this corresponds to whether the $G$-action is equivariant or not, and ultimately whether the phase space needs to be extended in order to accommodate the symmetry algebra. 

\myfig{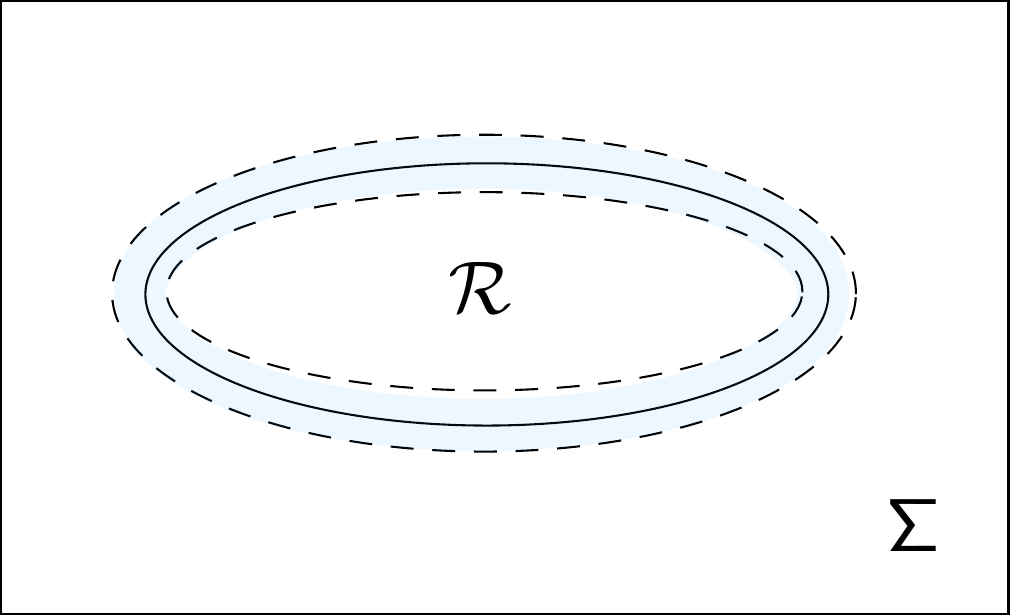}{5}{
The crossed product of the naive subregion algebra with the UCS may be understood as the algebra of ultra-local operators in $\mathcal{R}$ along with operators that have support in the tubular neighborhood of the entangling surface, an infinitesimal region generated by the UCS generators. 
}

The duality of these two procedures is realized (formally) via  geometric quantization. Given a (pre-) symplectic manifold, there is a mapping from its algebra of functions into an operator algebra that acts on a suitably defined Hilbert space. Thus, our aim is to show that the following diagram commutes:
\begin{equation} \label{commutative diagram}
\begin{tikzcd}
(X,\Omega) \arrow[r, "GQ"] \arrow[d, "EPS"]
& \mathcal{M} \arrow[d, "CP"] \\
(X_{ext},\Omega_{ext}) \arrow[r, "GQ"]
& \mathcal{R}(\mathcal{M},G)
\end{tikzcd}
\end{equation}
Here $GQ$ is geometric quantization, $EPS$ is the extension of the phase space, and $CP$ is the crossed product construction with $\mathcal{R}(\mathcal{M},G)$ denoting the crossed product algebra. In words: one can either start with the symplectic manifold $(X,\Omega)$, perform geometric quantization to realize the operator algebra $\mathcal{M}$ and then incorporate symmetries by performing the crossed product to obtain $\mathcal{R}(\mathcal{M},G)$; or one can start with the symplectic manifold $(X,\Omega)$, extend the phase space in order to accommodate the symmetry generators, and then perform geometric quantization resulting in the same algebra $\mathcal{R}(\mathcal{M},G)$. 

The main physical contribution of this note is to implement Takesaki's crossed product construction \cite{takesaki1973crossed} in the general context of the universal corner symmetry, and subsequently to show that the result can be interpreted as a quantization of the extended phase space formalism introduced in \cite{Klinger:2023qna}. We demonstrate how this crossed product covers the modular automorphism group for a wide class of subregions. What's more, we argue that crossing with the full UCS encodes the unitary flow of states produced by a continuous deformation to the geometry of the entangling surface between a subregion and its causal complement. In this way we recognize how, by combining non-extended field configurations with the data specifying how these degrees of freedom transform on spatial intersections, we realize an algebra which is semi-finite and related to the factorizable ``extended Hilbert space" introduced in \cite{Donnelly:2016auv}. It is worth noting that our construction is  independent of the AdS/CFT correspondence, and makes no reference to large $N$. In light of this fact, we argue that the extended phase space should be regarded as a universal regularization scheme for entanglement entropy divergences in arbitrary quantum field theories. 

This note is roughly divided into three major parts. In Section \ref{sec: Operator Algebras} we present the utility of the crossed product and the dual weight theorem in the analysis of type III von Neumann algebras. We begin with Section \ref{sec: CP} in which we recall the crossed product construction for a generic von Neumann algebra which admits a $G$-automorphism. In Section \ref{sec: typeIII} we address how the crossed product of a type III von Neumann algebra, $\mathcal{M}$, with its modular automorphism group generates a unique type II von Neumann algebra, $\mathcal{N}_{\mathcal{M}}$, representing $\mathcal{M}$.\footnote{In Appendix \ref{sec: mod} we review the role of weights in the study of von Neumann algebras, and introduce the modular automorphism group.} We state the dual weight theorem which allows for each weight on $\mathcal{M}$ to be uniquely associated with a weight on $\mathcal{N}_{\mathcal{M}}$. 
Moreover, we introduce an extension of the ordinary dual weight theorem which allows for states on $\mathcal{M}$ to be associated with states on $\mathcal{N}_{\mathcal{M}}$, provided one supplies the additional information associated with an embedding $\zeta: \mathcal{M} \hookrightarrow \mathcal{N}_{\mathcal{M}}$.
Because $\mathcal{N}_{\mathcal{M}}$ is a semi-finite algebra, by construction, the modified dual weight correspondence allows for states on $\mathcal{M}$ to be identified with density operators in $\mathcal{N}_{\mathcal{M}}$. This provides a fully algebraic prescription for assigning a meaningful entropy measure directly to states in a type III von Neumann algebra. In Section \ref{sec: extHilb} we compare the crossed product approach with the extended Hilbert space of \cite{Donnelly:2016auv}.

In Section \ref{sec: SymGeom} we motivate the correspondence \eqref{commutative diagram}. To begin, we review the construction of the prequantum Poisson algebra $\mathcal{M}_{pq}$ starting from a (pre-) symplectic manifold $(X,\Omega)$. In Section \ref{sec: Gaction} we study the case where $(X,\Omega)$ admits a $G$-action, and identify important analogies between this case and the set up of the crossed product. In Section \ref{sec: EPS} we introduce the extended phase space in complete generality starting with a (pre-) symplectic manifold $(X,\Omega)$ for which the aforementioned $G$-action is presymplectic. Finally, in Section \ref{sec: GQuant} we recall the procedure of geometric quantization, and establish the commutativity of \eqref{commutative diagram}. 

As we have introduced, the interesting use case of the crossed product and the extended phase space arises when one considers the physics of fields on a subregion of spacetime. Thus, we conclude in Section \ref{sec: example} by addressing this specific case, with the symmetry structure $G$ being supplied by the group of diffeomorphisms. Under suitable conditions, the modular automorphism group of the naive subregion algebra is realized as a subalgebra of the UCS. Thus, in such cases, one immediate consequence of extending the phase space to accommodate UCS charges is to implement Tomita-Takesaki theory for the subregion. In general, as we have alluded to, the complete UCS symmetry is necessary to allow for state changing deformations of the entangling surface. This observation ties a neat bow around the operator algebra picture and the corner symmetries by explicitly realizing how the addition of degrees of freedom governing the gluing of subregions regulates its entanglement singularity, thereby providing a physical interpretation for Tomita-Takesaki theory and the crossed product. It also suggests exciting next steps for studying quantum field theory in fully dynamical geometries.

\section{Operator Algebras and the Crossed Product} \label{sec: Operator Algebras}

In this Section we focus on the relevance of the crossed product construction in facilitating the study of type III von Neumann algebras. To begin, we implement the crossed product in great generality as the natural algebraic representation of a symmetry action. Our presentation follows closely with that of \cite{takesaki1973crossed}. Next, we discuss how the crossed product of a type III von Neumann algebra, $\mathcal{M}$, with its modular automorphism group furnishes a unique type II von Neumann algebra, $\mathcal{N}_{\mathcal{M}}$, representing it. We state the dual weight theorem which allows us to represent faithful weights on $\mathcal{M}$ as faithful weights on $\mathcal{N}_{\mathcal{M}}$, 
and introduce a modification which moreover ensures that states are mapped to states.
Because $\mathcal{N}_{\mathcal{M}}$ is a semi-finite algebra, this allows for the identification of faithful states on $\mathcal{M}$ with density operators in $\mathcal{N}_{\mathcal{M}}$ and subsequently for the computation of an associated entanglement entropy. 

\subsection{The Crossed Product Algebra} \label{sec: CP}

Let $\mathcal{M}$ denote a von Neumann algebra with involution $*$, and $\text{Aut}(\mathcal{M})$ denote the group of all $*$-preserving automorphisms of $\mathcal{M}$. Given a symmetry group $G$, a continuous action of $G$ on $\mathcal{M}$ is realized as a homomorphism $\alpha: G \rightarrow \text{Aut}(\mathcal{M})$ such that each $\alpha_g: \mathcal{M} \rightarrow \mathcal{M}$ is a continuous function in the strong operator topology. In a more pared down way of speaking, let us suppose that $\mathcal{M}$ is represented as $\rho:\mathcal{M} \to \mathcal{B}(\mathcal{H})$ --- that is, $\mathcal{M}$ is realized as a subalgebra of the bounded operators on a Hilbert space $\mathcal{H}$. In this case we can always construct a continuous action in terms of a unitary representation\footnote{Here, $\mathcal{U}(\mathcal{H})$ is the set of operators in $\mathcal{B}(\mathcal{H})$ that are unitary with respect to the inner product on $\mathcal{H}$.} $U: G \rightarrow \mathcal{U}(\mathcal{H})$ such that
\begin{equation} \label{Unitary auto}
\rho\circ\alpha_g(x) = U(g) \rho(x) U(g)^*, \qquad \forall x \in \mathcal{M}, g \in G. 
\end{equation}
From this perspective, we say that $\alpha_g$ is an \emph{inner} automorphism if $U(G) \subset \Im(\rho)$, otherwise it is an \emph{outer} automorphism. The triple $(\mathcal{M},G,\alpha)$ is called a \emph{covariant system} \cite{doplicher1966covariance}. 

Given a covariant system $(\mathcal{M},G,\alpha)$ we can  construct the crossed product of $\mathcal{M}$ by $G$ which we denote by $\mathcal{R}(\mathcal{M},G)$. The crossed product is realized as a subalgebra of the bounded operators on a related  Hilbert space that we will call $L^2(\mathcal{H};G,\mu)$; here $\mu$ is the left-invariant Haar measure on $G$ which we will use to write an inner product on $L^2(\mathcal{H};G,\mu)$. In particular, $L^2(\mathcal{H};G,\mu)$ can be regarded as the set of maps $\psi: G \rightarrow \mathcal{H}$ which have finite norm in the inner product $H: L^2(\mathcal{H};G,\mu) \times L^2(\mathcal{H};G,\mu) \rightarrow \mathbb{C}$ given by
\begin{equation}
    H(\xi,\eta) := \int_G \mu(g) h(\xi(g),\eta(g)),\qquad \forall \xi,\eta\in L^2(\mathcal{H};G,\mu).
\end{equation}
Here $h: \mathcal{H} \times \mathcal{H} \rightarrow \mathbb{C}$ is the inner product on $\mathcal{H}$. Acting on this Hilbert space we have the pair of representations
\begin{equation}
    \pi_\alpha: \mathcal{M} \rightarrow \mathcal{B}(L^2(\mathcal{H};G,\mu)), \qquad \lambda: G \rightarrow \mathcal{B}(L^2(\mathcal{H};G,\mu)),
\end{equation}
which are given explicitly by\footnote{$\pi_\alpha$ represents $\mathcal{M}$ in the sense that 
\[ 
(\pi_\alpha(xy)(\xi))(g)
=\rho\circ\alpha_g^{-1}(xy)(\xi(g))
=\rho\circ\alpha_g^{-1}(x)(\rho\circ\alpha_g^{-1}(y)(\xi(g)))
=\rho\circ\alpha_g^{-1}(x)((\pi_\alpha(y)(\xi))(g))
=(\pi_\alpha(x)\pi_\alpha(y)(\xi))(g)
\] and $\lambda$ represents $G$ in the sense that
\[ 
(\lambda(gk)(\xi))(h)
=\xi(k^{-1}g^{-1}h)
=(\lambda(k)(\xi))(g^{-1}h)
=(\lambda(g)\lambda(k)(\xi))(h)
.\]}
\begin{equation} \label{Crossed Product Rep}
    (\pi_\alpha(x)(\xi))(g) = \rho\circ\alpha^{-1}_g(x)(\xi(g)), \qquad (\lambda(g)(\xi))(h) = \xi(g^{-1}h),
\end{equation}
for each $\xi \in L^2(\mathcal{H};G,\mu),$ $g,h \in G,$ and $ x \in \mathcal{M}$. Equations \eqref{Crossed Product Rep} ensure that the automorphism is faithfully encoded in the space of bounded operators, i.e.,
\begin{equation} \label{Automorphism Condition}
    \lambda(g)\pi_\alpha(x) \lambda(g)^* = \pi_{\alpha} \circ \alpha_g(x),
\end{equation}
the analogue of \eqref{Unitary auto}.
The crossed product algebra is then equivalent to the von Neumann algebra generated by the union of $\pi_\alpha(\mathcal{M})$ and $\lambda(G)$, which is sometimes denoted by $\mathcal{R}(\mathcal{M},G) \equiv \pi_{\alpha}(\mathcal{M}) \rtimes \lambda(G)$.  

Although our construction has made use of the Hilbert space $\mathcal{H}$, the crossed product is independent of such a choice. In fact, the algebraic structure of $\mathcal{R}(\mathcal{M},G)$ is totally independent of a chosen representation, which motivates the notation $\mathcal{R}(\mathcal{M},G) = \mathcal{M} \rtimes_{\alpha} G$ where now we have dropped reference to the representations $\pi_{\alpha}$ and $\lambda$. In this respect, the crossed product is merely an algebra that includes as subalgebras the original algebra $\mathcal{M}$, and the group algebra $G$, and which allows for the direct action of the group $G$ on $\mathcal{M}$ in a way that preserves the group structure of $G$. In some sense, one should regard the crossed product as the minimal extension to the original algebra for which the automorphism is guaranteed to be inner. As a heuristic, the crossed product may be regarded as simply ``adding" the symmetry generators explicitly into the algebra, i.e., 
\beq \label{Closure of Crossed Product}
	\mathcal{R}(\mathcal{M},G) \sim \{\mathcal{M}, U(G)\}'',
\eeq
where $U(G)$ is the set of unitary operators carrying out the automorphism of $G$ on $\mathcal{M}$ in \eqref{Unitary auto}, and we have generated a von Neumann algebra by taking the double commutant. 

\subsection{The Dual Weight Theorem} \label{sec: typeIII}

As we have alluded to in the introduction, one of the primary uses of the crossed product construction is in the study of type III von Neumann algebras \cite{hiai2020concise,takesaki2003theory}. In \cite{takesaki1973crossed}, Takesaki proposed the crossed product of a type III von Neumann algebra $\mathcal{M}$ with its modular automorphism group as a unique type II representative of $\mathcal{M}$. This perspective has found renewed interest in recent physics literature as an approach to understanding the semi-finite operator algebra associated with a subregion in the context of the AdS/CFT correspondence \cite{Witten:2021unn}. In this section we provide an overview of the relationship between a type III algebra $\mathcal{M}$ and its crossed product by $\sigma: \mathbb{R} \times \mathcal{M} \rightarrow \mathcal{M}$ where $\sigma$ is the modular automorphism group.\footnote{In Appendix \ref{sec: mod} we provide an introduction to the modular automorphism group through the study of faithful, semi-finite, normal weights.} In particular, we state the dual weight theorem which allows for weights on the original algebra $\mathcal{M}$ to be uniquely identified with weights in the crossed product algebra $\mathcal{N}_{\mathcal{M}}$. This suggests a fully algebraic approach to assigning density operators and, by extension, entropy measurements to states in a type III von Neumann algebra. 

Let $\mathcal{N}_{\mathcal{M}} = \mathcal{M} \rtimes_{\sigma} \mathbb{R}$ denote the crossed product of $\mathcal{M}$ by its modular automorphism group. Denote by $\hat{\mathbb{R}}$ the Pontryagin dual group (or character group) of $\mathbb{R}$, i.e., $\hat{\mathbb{R}}$ consists of all maps from $\mathbb{R}$ into the circle group $U(1) \simeq S^1 \subset \mathbb{C}$: 
\beq
	\hat{\mathbb{R}} \equiv \{p: \mathbb{R} \rightarrow S^1\}.
\eeq
Explicitly, one may regard $\hat{\mathbb{R}}$ as a one-dimensional Abelian group, with the pairing of $\mathbb{R}$ and $\hat{\mathbb{R}}$ given by
\beq
	p_q(t) = e^{2\pi i q t}, \; q, t \in \mathbb{R}. 
\eeq 
There is a natural unitary representation of $\hat{\mathbb{R}}$ acting on the Hilbert space defined as in Section \ref{sec: CP}, $L^2(\mathcal{H};\mathbb{R},\mu)$:
\beq \label{Dual unitary}
	v: \hat{\mathbb{R}} \rightarrow \mathcal{U}(L^2(\mathcal{H};\mathbb{R},\mu)), \qquad (v(p)(\xi))(t) = \overline{p(t)} \xi(t), \; \xi \in L^2(\mathcal{H}; \mathbb{R},\mu).
\eeq
Notice that $v(p)$ only acts on the argument of $\xi$ and not on its image in the Hilbert space $\mathcal{H}$.\footnote{Identifying $L^2(\mathcal{H};\mathbb{R},\mu) \simeq \mathcal{H} \otimes L^2(\mathbb{R},\mu)$, we could write $v(p) = \mathbb{1} \otimes v_p$ where $(v_p(f))(t) = \overline{p(t)}f(t)$ for every $f \in L^2(\mathbb{R},\mu)$.}

Using the representation \eqref{Dual unitary}, we define an automorphism $\hat{\sigma}: \hat{\mathbb{R}} \times \mathcal{N}_{\mathcal{M}} \rightarrow \mathcal{N}_{\mathcal{M}}$ given explicitly by
\beq \label{Dual automorphism}
	\hat{\sigma}_p(\pi_{\alpha}(x)) = \pi_{\alpha}(x), \qquad \hat{\sigma}_p(\lambda(t)) = \overline{p(t)}\lambda(t), \;\; p \in \hat{\mathbb{R}}, t \in \mathbb{R}, x \in \mathcal{M}.
\eeq
The map $\hat{\sigma}$ is called the \emph{dual action}. The significance of the dual action is that it allows us to identify the original von Neumann algebra $\mathcal{M}$ as a subalgebra inside of the crossed product $\mathcal{N}_{\mathcal{M}}$. In particular, motivated by \eqref{Dual automorphism}, one can show that $\pi_{\alpha}(\mathcal{M})$ is the fixed point algebra of the dual action:
\beq
	\pi_{\alpha}(\mathcal{M}) = \{a \in \mathcal{N}_{\mathcal{M}} \; | \; \hat{\sigma}_p(a) = a, \; \forall p \in \hat{\mathbb{R}}\}.
\eeq
This implies that there exists a projection $\Pi: \mathcal{N}_{\mathcal{M}} \rightarrow \mathcal{M}$, from the crossed product algebra down to the fixed point set of the dual action. Formally, this projection map is a faithful, semi-finite, normal operator-valued weight from $\mathcal{N}_{\mathcal{M}}$ to $\mathcal{M}$. It has an explicit form in terms of the integral:
\beq \label{Pi map}
	\Pi(a) = \pi_{\alpha}^{-1}\left(\int_{\hat{\mathbb{R}}} dp \; \hat{\sigma}_p(a)\right)
\eeq
where $dp$ is the Haar measure on the dual group $\hat{\mathbb{R}}$. The triple $(\mathcal{N}_{\mathcal{M}},\mathbb{R},\hat{\sigma})$ form what is called the \emph{associated covariant system} of $\mathcal{M}$ \cite{takesaki2003theory}.

We are now prepared to state the dual weight theorem \cite{Haagerup1978I,Haagerup1978II,Digernes:414178}. Recall that a weight on a von Neumann algebra $\mathcal{M}$ is a linear map $\omega: \mathcal{M}_+ \rightarrow \mathbb{R}_+$.\footnote{Here $\mathcal{M}_+$ is the set of elements in $\mathcal{M}$ which have strictly non-negative spectra.} We denote the set of elements in $\mathcal{M}$ for which $\omega(x)$ is finite by
\beq
	\mathfrak{p}_{\omega} = \{ x \in \mathcal{M}_+ \; | \; \omega(x) < \infty \}.
\eeq
The weight is termed \emph{semi-finite} if the closure of $\mathfrak{p}_{\omega}$ in the weak operator topology is equivalent to $\mathcal{M}$ itself. The weight is termed \emph{faithful} if it is nonzero when acting upon any nonzero element in $\mathcal{M}$ i.e., $\omega(x) \neq 0$ provided $x \neq 0$. {Finally, the weight is termed \emph{normal} if it is continuous in the ultraweak operator topology.}  Weights are the operator algebraic generalization of density operators, and $\omega(x)$ may be interpreted as the expectation value of the operator $x$ in the ``state" $\omega$. This generalization is crucial to the study of type III von Neumann algebras since such algebras do not admit any bonafide density operators. However, in \cite{takesaki2003theory} it is proven that every von Neumann algebra (regardless of type) admits a faithful, semi-finite, normal weight. {In the following all weights are assumed to be faithful, semi-finite, and normal unless otherwise stated and so we will often omit these terms for brevity.}

The dual weight theorem states that, given a faithful, semi-finite, normal weight $\varphi$ on $\mathcal{M}$ there exists a faithful, semi-finite, normal weight $\tilde{\varphi}$ on $\mathcal{N}_{\mathcal{M}}$ such that
\beq \label{Dual Weight Theorem}
	\tilde{\varphi} = \varphi \circ \Pi. 
\eeq
This theorem is valid in the general context where $G$ is any locally compact group, however in the case that $G$ is abelian -- as is the case for the modular automorphism group -- it can moreover be shown that the correspondence $\varphi \rightarrow \tilde{\varphi}$ is unique. 

The significance of the dual weight theorem should be clear. Even if the algebra $\mathcal{M}$ is type III and therefore does not admit a trace or density operators, by construction the algebra $\mathcal{N}_{\mathcal{M}}$ is semi-finite. Thus, there exists a well-defined faithful, semi-finite, normal trace\footnote{In fact this trace is canonically defined starting from the modular automorphism group of any faithful weight $\varphi$ on the original algebra $\mathcal{M}$. See \cite{Haagerup1978I,Haagerup1978II,hiai2020concise,takesaki2003theory,haagerup1979lp} for details on the explicit formulation of this trace, and its properties.} $\tau: \mathcal{N}_{\mathcal{M}} \rightarrow \mathbb{C}$, and we can define density operators $\rho \in \mathcal{N}_{\mathcal{M}}$ as the set of elements which are positive, self-adjoint, and normalized under the trace $\tau$. 

Given a weight $\omega$ on $\mathcal{N}_{\mathcal{M}}$ we may therefore identify a positive, self-adjoint operator $\rho_{\omega} \in \mathcal{N}_{\mathcal{M}}$ such that
\beq
	\omega(a) = \tau(\rho_{\omega} a), \; a \in \mathcal{N}_{\mathcal{M}}.
\eeq
Notice, however, that we have stopped short of calling $\rho_{\omega}$ a density operator.\footnote{We are grateful to Jonathan Sorce for questions which motivated the extension of this section.} This is because for a generic weight the dual operator $\rho_{\omega}$, while positive and self-adjoint, may not be normalized. The normalization of $\rho_{\omega}$ corresponds to whether or not $\omega$ is a \emph{state}. A state is a weight for which $\varphi(\mathbb{1}) = 1$. If $\omega$ is a state, $\omega(\mathbb{1}) = \tau(\rho_{\omega}) = 1$, and thus $\rho_{\omega}$ can be called a bonafide density operator.

This brings us to an important point. Although the dual weight theorem maps weights on $\mathcal{M}$ to weights on $\mathcal{N}_{\mathcal{M}}$, it does not map \emph{states} to \emph{states}. The nexus of this observation is the presence of the non-normalizable Haar measure in \eqref{Pi map}. We can circumvent this difficulty by modifying the dual weight theorem slightly; in particular we can replace the operator-valued weight $\Pi$ with a generalized conditional expectation map.\footnote{For a brief review of generalized conditional expectations, see Appendix \ref{app: condexp}.} In \cite{accardi1982conditional} it is shown that, starting from a von Neumann algebra $A$ with a faithful, semi-finite, normal weight $\tilde{\varphi}$, and a von Neumann subalgebra $B \subseteq A$, one can always construct a unital, completely positive map $\zeta_*: A \rightarrow B$ such that
\beq \label{Generalized Conditional Expectation}
	\tilde{\varphi} = \varphi\rvert_B \circ \zeta_*.
\eeq
Here $\varphi\rvert_B$ is the restriction of the weight $\tilde{\varphi}$ to $B$. The map $\zeta_*$ is constructed as the bi-dual of a given embedding $\zeta: B \hookrightarrow A$. Returning to our specific case, we start with a weight $\varphi$ on $\mathcal{M} \subset \mathcal{N}_{\mathcal{M}}$ and, in principle, an embedding $\zeta: \mathcal{M} \hookrightarrow \mathcal{N}_{\mathcal{M}}$. Thus, we can ``invert" \eqref{Generalized Conditional Expectation} to deduce a weight $\tilde{\varphi}_{\zeta}$ on $\mathcal{N}_{\mathcal{M}}$ for which 
\beq
	\tilde{\varphi}_{\zeta} = \varphi \circ \zeta_*.
\eeq
We refer to $\tilde{\varphi}_{\zeta}$ as the dual weight of $\varphi$ relative to the embedding $\zeta$. Since $\zeta_*$ is a unital map,
\beq
	\tilde{\varphi}_{\zeta}(\mathbb{1}_{\mathcal{N}_{\mathcal{M}}}) = \varphi(\mathbb{1}_{\mathcal{M}}).
\eeq
Thus, if $\varphi$ is a state, $\tilde{\varphi}_{\zeta}$ will be a state as well. The cost of this modification is that the dual weight assignment is no longer unique, but rather depends on the choice of $\zeta$. 

The upshot of the modified dual weight construction is that each pair $(\varphi,\zeta)$, where $\varphi$ is a weight on $\mathcal{M}$ and $\zeta$ is an embedding of $\mathcal{M}$ in $\mathcal{N}_{\mathcal{M}}$, define a weight, $\tilde{\varphi}_{\zeta}$ on $\mathcal{N}_{\mathcal{M}}$. Moreover, the assignment is such that if $\varphi$ is a state on $\mathcal{M}$, $\tilde{\varphi}_{\zeta}$ will be a state on $\mathcal{N}_{\mathcal{M}}$. The latter property is possible because, in contrast to the original dual weight theorem, the modified version takes advantage of the additional degrees of freedom present in the crossed product in such a a way as to `smooth out' the weight $\tilde{\varphi}_{\zeta}$. To this point, we echo \cite{accardi1982conditional} in identifying the role played by the embedding and the generalized conditional expectation map dual to it; $\zeta_*$ locates, in a statistical sense, the algebra $\mathcal{M}$ inside of $\mathcal{N}_{\mathcal{M}}$. Here we are taking inspiration from the notion of a conditional expectation in the context of an Abelian operator algebra, which corresponds to an ordinary conditional probability distribution in the measure theoretic sense.

Using the modified dual weight theorem we can identify each faithful, semi-finite, normal state $\varphi$ on $\mathcal{M}$ with a density operator $\rho_{\tilde{\varphi}_{\zeta}} \in \mathcal{N}_{\mathcal{M}}$ by the composition of maps:
\beq
	\varphi \mapsto \tilde{\varphi}_{\zeta} \mapsto \rho_{\tilde{\varphi}_{\zeta}}.
\eeq
In this way we can assign a meaningful entropy directly to the state $\varphi$ modulo a choice of $\zeta$:\footnote{Qualitatively, one might like to think of the entropy $S(\varphi;\zeta)$ as arising from two contributions -- the first corresponding to the entropy of the state $\varphi$ and the second to the entropy of the conditional expectation $\zeta_*$, resulting in the sum
\beq \label{Generalized Entropy}
	S(\varphi;\zeta) \sim S(\varphi) + S(\zeta).
\eeq
The entropy on the left hand side is finite, while the entropy of the state $\varphi$ is divergent. Thus, the entropy of the conditional expectation must have a divergence that cancels with that of the state. The structure of \eqref{Generalized Entropy} as a sum, and the interpretation of canceling divergences is very reminiscent of the generalized entropy \cite{Bekenstein:1972tm, Bekenstein:1973ur, Bekenstein:1974ax}. In that case, $S(\varphi)$ would be interpreted as the entropy of quantum fields on a fixed gravitational background, and $S(\zeta)$ computes the area of an associated surface. This is related to observations made in \cite{Witten:2021unn, Chandrasekaran:2022eqq, Jensen:2023yxy, AliAhmad:2023etg} that the generalized entropy can be given a microscopic interpretation through the crossed product, in the semi-classical regime. We plan to formalize this analogy in future work.
}
\beq \label{Entropy in CP}
	S(\varphi; \zeta) \equiv -\tau\bigg(\rho_{\tilde{\varphi}_{\zeta}} \ln(\rho_{\tilde{\varphi}_{\zeta}})\bigg).
\eeq
{We should note that the entropy \eqref{Entropy in CP} is only defined up to a potentially divergent additive constant. This constant arises due to a multiplicative ambiguity in the definition of the trace on the crossed product algebra. However, this constant is state independent and therefore can be subtracted from the entropy with impunity \cite{Jensen:2023yxy,AliAhmad:2023etg}.}

\subsection{Comparison with Extended Hilbert Space} \label{sec: extHilb}

The idea behind the dual weight theorem is very evocative of the extended Hilbert space construction \cite{Donnelly:2016auv,Fliss:2017wop, Fliss:2020cos}. In that context, one begins with the Hilbert space of a Cauchy surface $\mathcal{H}_{\Sigma}$, and seeks to compute the entanglement entropy of a state $\ket{\Psi} \in \mathcal{H}_{\Sigma}$ when restricted to a subregion $\mathcal{R} \subset \Sigma$. Naively, one would compute such an entropy by first formulating the reduced density operator of the state in the subregion $\mathcal{R}$ by tracing out the complementary region $\mathcal{R}^c$. However, as we have mentioned, it is a well-known fact that the Hilbert space $\mathcal{H}_{\Sigma}$ does not possess a tensor factorization as $\mathcal{H}_{\Sigma} = \mathcal{H}_{\mathcal{R}} \otimes \mathcal{H}_{\mathcal{R}^c}$. Thus, the operation of a partial trace is not well-defined, and by extension neither is the reduced density operator. This fact is intimately related to the type III nature of the subregion operator algebra. 

In \cite{Donnelly:2016auv}, it was argued that the failure of the naive tensor factorization is a manifestation of entangling degrees of freedom which live on the interface between the subregion and its complement. Hereafter, we refer to this interface as the entangling surface, $S$. These degrees of freedom play the essential role of gluing the subregions together in a quantum mechanically consistent way. By \emph{gluing} we mean a methodology for identifying (non gauge invariant) fields at the interface between two subregions. This is closely related to the role played by transition functions in an atlas for a differentiable manifold. In fact, in \cite{Klinger:2023qna}, we demonstrated how these gluing degrees of freedom have an explicit representation in terms of Lie algebroid isomorphisms, a generalized notion of transition maps relating different choices of gauge and coordinates describing local fields. Gluing two subregions in a path integral context requires an insertion  identifying fields on either side of an entangling surface up to a Lie algebroid isomorphism (that is, up to a gauge transformation and diffeomorphism). 

Returning to \cite{Donnelly:2016auv}, the authors proposed that there exists an enlarged Hilbert space $\mathcal{H}_{\Sigma}^{ext}$ for which a kind of tensor factorization is restored:
\beq \label{Entangling product}
	\mathcal{H}^{ext}_{\Sigma} = \mathcal{H}_{\mathcal{R}} \otimes_{G_S} \mathcal{H}_{\mathcal{R}^c}.
\eeq
The tensor product here is the so-called entangling product, and $G_S$ is the symmetry group associated with the entangling surface, containing the operators which enforce the interface boundary conditions that allow for the gluing of subregions in the path integral. One should interpret \eqref{Entangling product} as stating that the total degrees of freedom accounted for in a complete tensor factorization correspond to the local degrees of freedom in $\mathcal{R}$, the local degrees of freedom in $\mathcal{R}^c$ and the entangling degrees of freedom which are shared between them. To compute the entanglement entropy of the state $\ket{\Psi} \in \mathcal{H}_{\Sigma}$ one uses the fact that the naive Hilbert space is a subset of the extended Hilbert space $\mathcal{H}_{\Sigma} \subset \mathcal{H}^{ext}_{\Sigma}$ to embed $\ket{\Psi} \in \mathcal{H}^{ext}_{\Sigma}$, and then uses the tensor factorization of the extended Hilbert space to facilitate a well-defined partial trace. 

In comparison, for a generic type III von Neumann algebra $\mathcal{M}$ one might wish to assign an entropy measurement to a weight $\varphi$. However, as a type III algebra possesses neither a faithful trace nor trace class operators, it is not possible to associate such a weight with a density operator. Thus, it seems that one can only compute the relative entropy between two weights, for example by using relative modular theory. As we have now presented, however, there is a way to assign an entropy measurement to the weight $\varphi$ by means of the crossed product construction and the modified dual weight theorem. Namely, there exists an extended operator algebra $\mathcal{N}_{\mathcal{M}} = \mathcal{M} \rtimes_{\sigma} \mathbb{R}$ which is semi-finite and for which $\mathcal{M}$ can be realized as a subalgebra. Thus, to compute the entropy of the weight $\varphi$ one passes to a dual representation $\tilde{\varphi}_{\zeta}$, and then uses the faithful trace on $\mathcal{N}_{\mathcal{M}}$ to compute its entropy. 

\section{The Crossed Product and the Extended Phase Space} \label{sec: SymGeom}

The upshot of Section \ref{sec: Operator Algebras} is that faithful, semi-finite, normal states on an operator algebra $\mathcal{M}$ \emph{always} have a representation as density operators in the extended algebra formed by taking the crossed product of $\mathcal{M}$ with its modular automorphism group. This provides a fully algebraic approach to the entanglement singularity problem in the context of subregion quantum field theory. Namely, one can associate a well-defined entropy measure to any weight in the subregion algebra simply by passing to its associated density operator in the crossed product algebra. What we would now like to pursue is a physically motivated reasoning behind the utility of the modular automorphism group and its associated crossed product. In particular, we will seek to demonstrate how the crossed product algebra relevant to subregion quantum field theories can be interpreted in terms of the same entangling degrees of freedom that rescued tensor factorization in the extended Hilbert space construction. 

To this end, we first aim to show how the extended phase space construction can be interpreted through the lens of the crossed product. As we have introduced in \cite{Klinger:2023qna}, when applied to field theories the extended phase space provides a framework for studying the role of gluing degrees of freedom at entangling surfaces using tools from symplectic geometry. In that context, the non-factorizability problem manifests itself in the non-integrablility of charges associated with gauge symmetries in the presence of corners, i.e., co-dimension-$2$ submanifolds of spacetime. This non-integrability can be eliminated by extending the usual phase space of the theory (that is the phase space of gauge fixed field configurations) to include new degrees of freedom that explicitly correspond to transition functions, either for coordinate charts or choices of gauge, at the intersection between subregions \cite{Ciambelli:2021vnn, Ciambelli:2021nmv,Klinger:2023qna}. The extension of the phase space should be regarded as a prequantum analog of the extension of the Hilbert space advocated for in \cite{Donnelly:2016auv}. To bring this correspondence to the operator algebra level, we now aim to demonstrate how the crossed product provides an interpretation for the geometric quantization of the extended phase space. In pursuit of this fact, we will first motivate a more general correspondence between the extended phase space of generic symplectic geometries and the crossed product of generic von Neumann algebras. To orient the reader, we provide a basic blueprint of this correspondence now. 

Recall that a (pre-) symplectic manifold $(X,\Omega)$ gives rise to an algebra of functions called the Poisson algebra with a composition law endowed by the symplectic form. In physical contexts, the Poisson algebra may be regarded as the prequantum algebra, in the sense that a consistent quantization procedure lifts the algebra of functions into an operator algebra acting on a suitable Hilbert space. Symmetries are encoded in a symplectic geometry by means of a group action $\varphi: G \times X \rightarrow X$ which preserves the symplectic form, i.e., $\varphi_g^* \Omega = \Omega$ for all $g \in G$. Such an action therefore preserves the Poisson algebra, and may be regarded as an automorphism of the prequantum algebra in the sense described in Section \ref{sec: Operator Algebras}. 

At this stage, one may ask whether the infinitesimal action of the group $G$ on $(X,\Omega)$ can be represented by a family of functions living inside of the Poisson algebra itself. This is analogous to the question of whether an automorphism is inner or outer in the crossed product construction. In the symplectic context, this question is equivalent to the question of whether the group action $\varphi$ is equivariant or not. Recall that an equivariant group action on a symplectic manifold is one for which there exists an algebra homomorphism $\Phi: \mathfrak{g} \rightarrow \Omega^0(X)$ representing each generator of the Lie algebra ($G = \text{exp}(\mathfrak{g})$) as a function in the Poisson algebra. The fact that $\Phi$ is a homomorphism means that the Lie bracket is faithfully encoded in the Poisson algebra. In the case that $\varphi$ is equivariant, we therefore say that the automorphism generated by it is inner. Otherwise, we refer to the automorphism as  outer. 

As we have seen, the crossed product construction provides a method for extending an algebra in such a way as to ensure that a given automorphism is inner. Similarly, we introduce a general procedure for constructing an extended (pre-) symplectic manifold $(X_{ext},\Omega_{ext})$ for which any symmetry action can be lifted to an equivariant action. This is the extended phase space. At the level of the Poisson algebra, the extended Poisson algebra may be interpreted as a pre-crossed product in the sense that it is algebraically equivalent to the crossed product of the original Poisson algebra with the algebra of the symmetry group. Once an appropriate quantization procedure is implemented, the pre-crossed product becomes the crossed product of the quantized non-extended Poisson algebra and the symmetry group. This is our desired correspondence. 

\subsection{Preliminaries} \label{sec: SymGeomPrelims}

Let $X$ be a topological space. In physical contexts, we may take $X$ to be a suitable space of fields; for example, in the case of a gauge theory $X$ might consist of the components of gauge and matter fields. This space, together with 
a closed\footnote{Here and in the following we use the standard notation $Z^p(X) = \text{ker}(d_p)$ and $B^p(X) = \text{im}(d_{p-1})$ to refer to closed and exact $p$-forms on $X$. We use $\Omega^p(X)$ to refer to set of all $p$-forms. Finally $H^p(X) = Z^p(X)/B^p(X)$ is the $p^{th}$ cohomology class of $X$.} two-form $\Omega \in Z^2(X)$ on $X$ for which the map
\beq
	\tilde{\Omega}: TX \rightarrow T^*X; \qquad \un{V} \mapsto \tilde{\Omega}(\un{V}) = i_{\un{V}} \Omega,
\eeq
has constant rank, defines  a (pre-) symplectic manifold, denoted $(X,\Omega)$. Moreover, if $\tilde{\Omega}$ is an isomorphism, $(X,\Omega)$ is termed a symplectic manifold. Since $\Omega$ is a closed form, it is also at least locally exact. Thus (locally) there exists $\Theta \in \Omega^1(X)$ such that
\beq
	\Omega = \td \Theta.
\eeq
with $\td$ the de Rham differential on $X$. 
We refer to $\Theta$ as the presymplectic potential, and it will play an important role in understanding the construction of the extended phase space.  

A vector field $\un{V} \in TX$ is called \emph{presymplectic}\footnote{Naming conventions differ, and sometimes such a vector field is called symplectic.} if the form $i_{\un{V}}\Omega \in Z^1(X)$ is closed. We denote the set of presymplectic vector fields by $\text{Sym}(X)$. Notice that, given $\un{V} \in \text{Sym}(X)$, we have
\beq \label{presymplectic vector fields}
	\mathcal{L}_{\un{V}} \Omega = \td i_{\un{V}} \Omega + i_{\un{V}} \td \Omega = 0.
\eeq
Thus, \eqref{presymplectic vector fields} is an alternative characterization of the presymplectic vector fields. If, beyond being closed, $i_{\un{V}}\Omega \in B^1(X)$ is \emph{exact}, $\un{V}$ is called \emph{Hamiltonian}. We denote the set of Hamiltonian vector fields by $\text{Ham}(X)$ and note that $\text{Ham}(X) \subseteq \text{Sym}(X)$ for the same reason that $B^1(X) \subseteq Z^1(X)$. Whether or not these spaces coincide is clearly a question of the cohomology of $X$, in particular it is a characterization of $H^1(X)$. 

We denote by $\Omega^0(X)$ the space of functions on $X$. As we will now demonstrate the symplectic structure of $X$ promotes $\Omega^0(X)$ into an algebra, the Poisson algebra. In particular, the de Rham differential provides a map $\td: \Omega^0(X) \rightarrow B^1(X) \simeq \text{Ham}(X)$. We denote by $\un{V}_\psi \in \text{Ham}(X)$ the Hamiltonian vector field solving\footnote{Note that $\un{V}_\psi$ is only unique in the case that $\Omega$ is symplectic. Otherwise, we would  always be able to add to $\un{V}_\psi$ any vector field in the kernel of $\tilde{\Omega}$. Nevertheless, this ambiguity does not affect the Poisson bracket and so we can ignore it for present purposes.}
\beq
	\td\psi + i_{\un{V}_\psi} \Omega = 0,\qquad \psi\in\Omega^0(X).
\eeq
Given the Hamiltonian pair $(\psi,\un{V}_\psi)$ one defines the Poisson bracket on $\Omega^0(X)$:
\beq \label{Poisson Algebra}
	\{\psi,\phi\} \equiv \Omega(\un{V}_\psi, \un{V}_\phi).
\eeq
We regard the Poisson algebra as defining the prequantum operator algebra $\mathcal{M}_{pq}\subseteq \text{Der}(\Omega^0(X))$ which acts on the space of functions $\Omega^0(X)$ via \eqref{Poisson Algebra}. 

\subsection{G-actions on symplectic manifolds} \label{sec: Gaction}

Let us now consider the case that $(X,\Omega)$ hosts a $G$-action, $\varphi: G \times X \rightarrow X$. The $G$-action descends to the infinitesimal level by means of the homomorphism $\xi: \mathfrak{g} \rightarrow TX$ such that
\beq
	\xi_{\un{\mu}} = (\varphi_{\text{exp}(t\un{\mu})})_* \frac{d}{dt}, \qquad \un{\mu} \in \mathfrak{g}.
\eeq
Given a $G$-action on a (pre-) symplectic manifold it is conventional to classify the action with the following designations. Firstly, the $G$-action is termed \emph{presymplectic} if $\xi_{\un{\mu}} \in \text{Sym}(X)$ for each $\un{\mu} \in \mathfrak{g}$. That is
\beq \label{Symplectomorphism}
	\mathcal{L}_{\xi_{\un{\mu}}}\Omega = \td i_{\xi_{\un{\mu}}} \Omega = 0, \; \forall \un{\mu} \in \mathfrak{g}.
\eeq
At the level of the group action, this implies that $\varphi_g^*\Omega = \Omega$ for all $g \in G$. In other words, a presymplectic $G$-action assigns to each $g \in G$ an automorphism of $\mathcal{M}_{pq}$, since the (pre-) symplectic form and hence by extension the Poisson algebra of functions \eqref{Poisson Algebra} is preserved by this action. 

Next, the $G$-action is said to be \emph{Hamiltonian} if $\xi_{\un{\mu}} \in \text{Ham}(X)$ for each $\un{\mu} \in \mathfrak{g}$. As we have already addressed $\text{Ham}(X) \subseteq \text{Sym}(X)$, and thus a $G$-action which is Hamiltonian is automatically presymplectic, although the converse is not immediately true. Given a Hamiltonian $G$-action on $(X,\Omega)$, each $\un{\mu} \in \mathfrak{g}$ can be associated with an element $\Phi_{\un{\mu}} \in \Omega^0(X)$ such that
\beq \label{Hamiltonian}
	i_{\xi_{\un{\mu}}}\Omega + \td \Phi_{\un{\mu}} = 0.
\eeq
We regard this as defining a map $\Phi: \mathfrak{g} \rightarrow \Omega^0(X)$ which is referred to as the \emph{moment map}. 

Finally, the $G$-action is said to be \emph{equivariant} if the moment map $\Phi$ is a morphism of the Lie algebra bracket and the Poisson bracket, that is if\footnote{Alternatively, by viewing $\Phi: X \rightarrow \mathfrak{g}^*$ we can state \eqref{Equivariance} as the condition that $\Phi$ must be an equivariant map. That is
\beq
	\mathcal{L}_{\xi_{\un{\mu}}} \Phi(\un{\nu}) = \Phi([\un{\mu},\un{\nu}]) = \text{ad}^{*}_{\un{\mu}}(\Phi)(\un{\nu}).
\eeq
Here $\text{ad}^{*}: \mathfrak{g} \times \mathfrak{g}^* \rightarrow \mathfrak{g}^*$ is the co-adjoint action of $\mathfrak{g}$ on $\mathfrak{g}^*$. 
}
\beq \label{Equivariance}
	R^{\Phi}(\un{\mu},\un{\nu}) = \{\Phi_{\un{\mu}},\Phi_{\un{\nu}}\} - \Phi_{[\un{\mu},\un{\nu}]} = 0.
\eeq
Again, notice that \eqref{Equivariance} assumes the existence of a moment map, and thus equivariance is a more restrictive condition than Hamiltonian. 

Let us now introduce some nomenclature inspired by the crossed product. If the $G$-action $\varphi$ on $(X,\Omega)$ is presymplectic, we will say that $\varphi$ generates an automorphism of the prequantum algebra $\mathcal{M}_{pq}$. Indeed, this is merely the definition of a presymplectic action which preserves the symplectic form and by extension the Poisson bracket. If the $G$-action is moreover equivariant (and thus also Hamiltonian), we will say that the automorphism generated by $G$ is \emph{inner}. Conversely, if the $G$-action is not equivariant (even if it is Hamiltonian) we say that the automorphism generated by $G$ is \emph{outer}. The reasoning behind this terminology is as follows.

Suppose that the $G$-action on $(X,\Omega)$ is equivariant. Then, the full algebra of symmetry generators are faithfully encoded in $\mathcal{M}_{pq}$. To see why this is the case, let us first note that the existence of a moment map implies $\mathfrak{g}$ can be made to act on $\Omega^0(X)$ through the Poisson bracket \eqref{Poisson Algebra}. We denote this action by $\lambda: \mathfrak{g} \rightarrow \text{Der}(\Omega^0(X))$, with
\beq \label{Adjoint rep of sym}
	\lambda_{\un{\mu}}(f) = \{\Phi_{\un{\mu}},f\}. 
\eeq
Next, let us consider the question of whether \eqref{Adjoint rep of sym} defines a morphism of the brackets of $\mathcal{M}_{pq}$ and $\mathfrak{g}$ or not. We compute
\begin{flalign} \label{der bracket}
	[\lambda_{\un{\mu}},\lambda_{\un{\nu}}]_{\mathcal{M}_{pq}}(f) &= \{\Phi_{\un{\mu}},\{\Phi_{\un{\nu}},f\}\} - \{\Phi_{\un{\nu}},\{\Phi_{\un{\mu}},f\}\} \\
	&= \{\{\Phi_{\un{\mu}},\Phi_{\un{\nu}}\},f\},
\end{flalign}
where here we have used the antisymmetry of the Poisson bracket and the Jacobi identity. Combining \eqref{der bracket} and \eqref{Equivariance} we find:
\beq \label{Lambda curvature}
	R^{\lambda}(\un{\mu},\un{\nu}) := [\lambda_{\un{\mu}},\lambda_{\un{\nu}}]_{\mathcal{M}_{pq}} - \lambda_{[\un{\mu},\un{\nu}]} = \lambda_{R^{\Phi}(\un{\mu},\un{\nu})}  
\eeq
The action \eqref{Adjoint rep of sym} is a bracket homomorphism if and only if \eqref{Lambda curvature} vanishes, which, as we saw, will be the case provided the $G$-action is equivariant. In other words, given an equivariant $G$-action
\beq \label{Inner presymplectic}
	\text{im}(\lambda) \simeq \mathfrak{g} \subset \mathcal{M}_{pq}.
\eeq
This is analogous to the idea of an inner automorphism group in the context of the crossed product, hence the terminology used here.\footnote{In the context of symplectic geometry it is more natural to work at the level of the algebra generating the group $G$. However, for all cases we will be interested in we will be able to exponentiate $\mathfrak{g}$ and thus statements like \eqref{Inner presymplectic} can be carried into analogous statements about the group.}

We can follow the same logic as above in the case that the $G$-action is Hamiltonian, up to the vanishing of \eqref{Lambda curvature}. If the action is not equivariant, \eqref{Lambda curvature} \emph{will not vanish identically}. Thus, although each $\un{\mu} \in \mathfrak{g}$ can be identified with an element of $\Omega^0(X)$ via the moment map, it is \emph{not true} that the image of $\lambda$ is contained in the prequantum operator algebra $\mathcal{M}_{pq}$ because the Lie algebra of symmetry generators is not faithfully encoded. This fact is even more clear in the case that the $G$-action is merely presymplectic, whereupon we cannot even identify elements of $\mathfrak{g}$ with functions in $\Omega^0(X)$. Hence, in these cases we refer to the $G$ action as generating an outer automorphism for the same reason as in the crossed product.

\subsection{The Extended Phase Space} \label{sec: EPS}

We are now prepared to construct the extended phase space and observe its correspondence with the crossed product algebra. Our starting point is an automorphism of $(X,\Omega)$, meaning we have a presymplectic $G$-action as described above. The goal of the extended phase space is to identify a (pre-) symplectic structure $(X_{ext},\Omega_{ext})$ which can be constructed from $(X,\Omega)$ and $G$ only, and for which the presymplectic $G$-action on $(X,\Omega)$ can be lifted to an equivariant action. Provided such a space exists, the extended $G$-automorphism on $(X_{ext},\Omega_{ext})$ becomes an inner automorphism with the symmetry generators contained explicitly in the extended prequantum algebra. We denote the prequantum algebra of extended phase space of $(X,\Omega)$ by $G$ as $EPS(\mathcal{M}_{pq},G)$.\footnote{Note, in the case that the $G$-automorphism is already inner on $(X,\Omega)$, the extended phase space coincides with the original phase space. In other words, for an inner automorphism $EPS(\mathcal{M}_{pq},G) = \mathcal{M}_{pq}$. This is consistent with the fact that the crossed product of an algebra with an inner automorphism group does not change the algebra.} In this Section we make use of Atiyah Lie algebroids which are a very natural tool for analyzing the extended phase space. The unfamiliar reader should study Section 3 of \cite{Klinger:2023qna} before reading this Section for prerequisite analysis of the salient features of Lie algebroids used herein.

To begin, we define the extended phase space $X_{ext} = X \times G$. Formally, this characterization is only true locally. More rigorously, we should identify $X_{ext}$ as being a principal $G$-bundle over $X$. The structure maps of this principal bundle are given by
\beq \label{structure maps of Xext}
	\varphi: X_{ext} \rightarrow X, \; (g,x) \rightarrow \varphi_g(x), \qquad R: G \times X_{ext} \rightarrow X_{ext}, \; (g,x) \mapsto R_h(g,x) = (gh, \varphi_{h^{-1}}(x)),
\eeq
which make use of the $G$-action on $X$ and the right action of $G$ on itself. It is easy to see that $\varphi \circ R_h(g,x) = \varphi(gh,\varphi_{h^{-1}}(x)) = \varphi_g(x)$, and hence these maps possess the compatibility necessary to ensure that $X_{ext}$ is a principal bundle. 

Having extended the phase space $X \mapsto X_{ext}$, we must also perform an analogous extension of the symplectic form $\Omega \mapsto \Omega_{ext}$ in order to ensure that $(X_{ext},\Omega_{ext})$ can properly be interpreted as a symplectic structure. Rather than work with the tangent space to $X_{ext}$ as a model for the symplectic vector space, we will construct the Atiyah Lie algebroid associated with $TX_{ext}$. Formally, this means that we will work in the quotient space $A = TX_{ext}/G$, where the quotient is taken with respect to the right action of $G$ \cite{mackenzie2005general,crainic2003integrability}. The utility of performing this quotient is that it transforms $TX_{ext}$ from a vector bundle over $X_{ext}$ to a vector bundle over $X$. This will allow us to more clearly relate the extended symplectic structure to the non-extended symplectic structure. We refer to $A$ as the configuration algebroid.\footnote{For a further selection of recent papers using Lie algebroids in physics contexts see \cite{Klinger:2023qna,Jia:2023tki,Fournel:2012uv,Ciambelli:2021ujl,Blohmann:2010jd,LAZZARINI2012387,Carow-Watamura:2016lob,Kotov:2016lpx,ATTARD2020103541,Strobl:2004im,BOJOWALD2005400,Mayer:2009wf}.} 

The configuration algebroid inherits its structure maps from \eqref{structure maps of Xext}, resulting in the bundle maps
\beq
	\rho: A \rightarrow TX, \qquad j: L \rightarrow A.
\eeq
Here $L$ is a vector bundle over $X$ with standard fiber equal to $\mathfrak{g}$. Hence, we may identify  sections of $L$ with $X$-dependent symmetry generators. The maps $\rho$ and $j$ are bracket morphisms. We also introduce a Lie algebroid connection through the pair of maps
\beq
	\sigma: TX \rightarrow A, \qquad \omega: A \rightarrow L.
\eeq
The failure of these maps to be morphisms corresponds to the curvature of the horizontal distribution $H = \text{im}(\sigma) = \text{ker}(\omega) \subset A$. 

To promote differential forms on $TX$ into differential forms on $A$ we must pull back by the map $\rho$. In particular, we define the extended presymplectic potential:
\beq
	\Theta_{ext} = \rho^* \Theta \in \Omega^1(H;E).
\eeq
Here we have distinguished that $\Theta_{ext}$ lives in a non-trivial representation of the structure group, which we have labeled by the vector bundle $E$. The reader may wonder why this is the case, since \emph{a priori} the presymplectic potential is just a differential form with values in functions on $X$ -- which one typically regards as the trivial representation. However, if $\Theta$ \emph{were} in the trivial representation, the automorphism generated by $G$ on $(X,\Omega)$ would have to be inner. Indeed, let us note that the invariance of $\Theta$ under the $G$ action:
\beq \label{Theta invariance}
	\varphi_g^* \Theta = \Theta, \; \forall g \in G,
\eeq
is a sufficient (albeit perhaps not necessary) condition for the $G$-action to be equivariant. Equation \eqref{Theta invariance} implies:
\beq \label{Lie Theta = 0}
	\mathcal{L}_{\xi_{\un{\mu}}} \Theta = 0.
\eeq
If \eqref{Lie Theta = 0} held, we could write
\beq
	\td i_{\xi_{\un{\mu}}}\Theta + i_{\xi_{\un{\mu}}}\Omega = 0,
\eeq
hence identifying $\Phi_{\un{\mu}} = i_{\xi_{\un{\mu}}}\Theta$ as a moment map, and ensuring that the $G$ action is Hamiltonian. What's more, \eqref{Lie Theta = 0} implies that for all $\un{V} \in TX$
\begin{flalign} 
	0 &= i_{\un{V}}(\td \Phi_{\un{\mu}} + i_{\xi_{\un{\mu}}}\Omega) \\
	&= \mathcal{L}_{\un{V}} \Phi_{\un{\mu}} + i_{\un{V}}i_{\xi_{\un{\mu}}}\Omega.
\end{flalign} 
Specializing to the case that $\un{V} = \xi_{\un{\nu}} \in TX$, and using \eqref{Poisson Algebra} we therefore find
\begin{flalign} \label{equivariance 1}
	0 &= \mathcal{L}_{\xi_{\un{\nu}}} \Phi_{\un{\mu}} + \{\Phi_{\un{\mu}},\Phi_{\un{\nu}}\}
\end{flalign}
Finally, we can write
\begin{flalign} \label{equivariance 2}
	\mathcal{L}_{\xi_{\un{\nu}}} \Phi_{\un{\mu}} &= \mathcal{L}_{\xi_{\un{\nu}}} i_{\xi_{\un{\mu}}}\Theta = i_{\xi_{\un{\mu}}} \mathcal{L}_{\xi_{\un{\nu}}} \Theta + i_{[\xi_{\un{\nu}},\xi_{\un{\mu}}]}\Theta \\
	&= -i_{[\xi_{\un{\mu}},\xi_{\un{\nu}}]}\Theta = -i_{\xi_{[\un{\mu},\un{\nu}]}}\Theta = -\Phi_{[\un{\mu},\un{\nu}]}.
\end{flalign}
Here we have used \eqref{Lie Theta = 0} along with the fact that $\xi$ is a bracket homomorphism. Combining \eqref{equivariance 1} and \eqref{equivariance 2} we find
\beq
	0 = -\Phi_{[\un{\mu},\un{\nu}]} + \{\Phi_{\un{\mu}},\Phi_{\un{\nu}}\},
\eeq
which allows us to conclude that the $G$ action is also equivariant. Thus, in the case that $\Theta$ is invariant under the action of $G$, i.e., $\varphi_g^*\Theta = \Theta$, the $G$-action is automatically equivariant, and represents an inner automorphism. 

In the case that the $G$-automorphism is inner, no extension is necessary. So, it remains only to consider the case where \eqref{Lie Theta = 0} does not hold. Instead, let us assume that
\beq \label{Covariance}
	i_{\un{Y}}\mathcal{L}_{\xi_{\un{\mu}}} \Theta = v_E(\un{\mu})(i_{\un{Y}}\Theta),
\eeq
or in other words that the components of $\Theta$ transform under an appropriate representation of $L$, given here by $v_E: L \rightarrow \text{End}(E)$. We will now show that \eqref{Covariance} is a sufficient condition for the $G$-action on the extended phase space $X_{ext}$ to be equivariant. 

First, we define the extended (pre-) symplectic form
\beq \label{extended sym form}
	\Omega_{ext} = \hat\td \Theta_{ext}.
\eeq
Here $\hat\td$ is the nilpotent coboundary operator which is naturally defined on $\Omega(A;E)$ (the exterior algebra of $A$ with values in sections of the associated bundle $E$), meaning $\hat\td\Omega_{ext} = \hat\td^2 \Theta_{ext} = 0$. Thus, $\Omega_{ext}$ defines a bonafide presymplectic form on $X_{ext}$. The definition \eqref{extended sym form} has the benefit that, when restricted to horizontal vectors $\sigma(\un{X}),\sigma(\un{Y}) \in H$ the extended symplectic form reduces to the non-extended form
\beq
	\hat{i}_{\sigma(\un{Y})}\hat{i}_{\sigma(\un{X})}\Omega_{ext} = \hat{i}_{\sigma(\un{Y})}\hat{i}_{\sigma(\un{X})} \hat\td \rho^*\Theta = \hat{i}_{\sigma(\un{Y})}\hat{i}_{\sigma(\un{X})} \rho^*\td \Theta = i_{\un{Y}}i_{\un{X}}\Omega.
\eeq
In this series of equalities we have used the fact that $\rho: A \rightarrow TX$ is a chain map when restricted to the horizontal sub-bundle of $A$.\footnote{Note, therefore, that this argument does not carry over to the case of vertical sections of $A$ -- indeed, this is where the extension lies.} 

With \eqref{extended sym form} in hand, we will now show that there exists a map $\Xi: L \rightarrow A$ for which 
\beq
	\hat{\mathcal{L}}_{\Xi_{\un{\mu}}}\Theta_{ext} = 0, \forall \un{\mu} \in L,
\eeq
provided \eqref{Covariance} holds. We can compute\footnote{Again, for details on the definition of the coboundary operator $\hatd$ and the associated Lie algebroid representation $\phi_E$ we refer the reader to \cite{Klinger:2023qna,Jia:2023tki,Ciambelli:2021ujl}.}
\begin{flalign}
	\hat{i}_{\un{\mathfrak{Y}}} \hat{\mathcal{L}}_{\Xi_{\un{\mu}}}\Theta_{ext} &= \hat{i}_{\Xi_{\un{\mu}}} \hat\td \hat{i}_{\un{\mathfrak{Y}}} \Theta_{ext} - \hat{i}_{[\Xi_{\un{\mu}},\un{\mathfrak{Y}}]_A} \Theta_{ext} \\
	&= \phi_E(\Xi_{\un{\mu}})(i_{\rho(\un{\mathfrak{Y}})}\Theta) - i_{\rho([\Xi_{\un{\mu}},\un{\mathfrak{Y}}]_A)}\Theta \\
	&= \mathcal{L}_{\rho(\Xi_{\un{\mu}})}(i_{\rho(\un{\mathfrak{Y}})}\Theta) + v_E \circ \omega(\Xi_{\un{\mu}})(i_{\rho(\un{\mathfrak{Y}})}\Theta) - i_{[\rho(\Xi_{\un{\mu}}),\rho(\un{\mathfrak{Y}})]}\Theta \\
	&= i_{\rho(\un{\mathfrak{Y}})}\mathcal{L}_{\rho(\Xi_{\un{\mu}})}\Theta + v_E \circ \omega(\Xi_{\un{\mu}})(i_{\rho(\un{\mathfrak{Y}})}\Theta)
\end{flalign}
In the first line we have used the Kozsul formula to compute the Lie derivative. In the second line we used the fact that $\Theta_{ext} = \rho^*\Theta$, and the definition of $\hat\td$ in terms of the morphism $\phi_E$. In the third line, we used the fact that $\phi_E(\un{\mathfrak{X}})(\un{\psi}) = \mathcal{L}_{\rho(\un{\mathfrak{X}})}\un{\psi} + v_E\circ \omega(\un{\mathfrak{X}})(\un{\psi})$ for any zero form $\un{\psi}$ in a representation space $E$. Finally, in the last line we used the Cartan algebra. Now, let us write $\Xi_{\un{\mu}} = \sigma(\xi_{\un{\mu}}) \oplus j(\un{\mu})$. Then, we see that
\beq \label{Lie der extended}
	\hat{i}_{\un{\mathfrak{Y}}}\hat{\mathcal{L}}_{\Xi_{\un{\mu}}}\Theta_{ext} = i_{\rho(\un{\mathfrak{Y}})}\mathcal{L}_{\xi_{\un{\mu}}}\Theta - v_E(\un{\mu})(i_{\rho(\un{\mathfrak{Y}})}\Theta),
\eeq
which will be equal to zero precisely if \eqref{Covariance} holds.

The combination of equations \eqref{Covariance} and \eqref{Lie der extended} ensures that the $G$-action generated by $\Xi_{\un{\mu}}$ on the extended symplectic space $(X_{ext},\Omega_{ext})$ is Hamiltonian, and identifies
\beq \label{Extended moment map}
	\Phi_{\un{\mu}} = \hat{i}_{\Xi_{\un{\mu}}}\Theta_{ext} = i_{\xi_{\un{\mu}}}\Theta
\eeq
as the associated moment map. The definition of the moment map \eqref{Extended moment map} depends only on $\xi_{\un{\mu}}$ and the non-extended symplectic potential, this is crucial in ensuring equivariance. Indeed, one might have been worried that the appearance of the connection in the definition of the map $\Xi$ could have signaled a failing of the calculation in \eqref{equivariance 2}, which required that $\xi$ was a bracket homomorphism. However, because the moment map only depends on $\xi$, it is straightforward to show that equivariance also follows from the vanishing of $\hat{\mathcal{L}}_{\Xi_{\un{\mu}}}\Theta_{ext}$. In particular we can compute:
\begin{flalign}
	\hat{\mathcal{L}}_{\Xi_{\un{\nu}}} \Phi_{\un{\mu}} &= \hat{\mathcal{L}}_{\Xi_{\un{\nu}}} \hat{i}_{\Xi_{\un{\mu}}}\Theta_{ext} = \hat{i}_{\Xi_{\un{\mu}}} \hat{\mathcal{L}}_{\Xi_{\un{\nu}}}\Theta_{ext} + \hat{i}_{[\Xi_{\un{\nu}},\Xi_{\un{\mu}}]}\Theta_{ext} \\
	&= i_{\rho([\Xi_{\un{\nu}},\Xi_{\un{\mu}}])}\Theta = i_{[\rho(\Xi_{\un{\nu}}),\rho(\Xi_{\un{\mu}})]}\Theta = i_{[\xi_{\un{\nu}},\xi_{\un{\mu}}]}\Theta = i_{\xi_{[\un{\nu},\un{\mu}]}}\Theta = \Phi_{[\un{\nu},\un{\mu}]},
\end{flalign} 
and by extension:
\beq
	0 = \hat{i}_{\Xi_{\un{\nu}}} \hat{\mathcal{L}}_{\Xi_{\un{\mu}}}\Theta_{ext} = \hat{\mathcal{L}}_{\Xi_{\un{\nu}}}\Phi_{\un{\mu}} + \{\Phi_{\un{\mu}},\Phi_{\un{\nu}}\}_{ext} = -\Phi_{[\un{\mu},\un{\nu}]} + \{\Phi_{\un{\mu}},\Phi_{\un{\nu}}\}_{ext}.
\eeq
We therefore conclude that provided \eqref{Covariance} holds we are guaranteed that the extended phase space $(X_{ext},\Omega_{ext})$ admits an equivariant $G$ action. 

At last, we can now compare the prequantum algebra of the non-extended phase space, $\mathcal{M}_{pq}$, with that of the extended phase space, $EPS(\mathcal{M}_{pq},G)$. The extended prequantum algebra consists of all the elements of $\mathcal{M}_{pq}$ along with the addition of the maps $\Phi_{\un{\mu}}$ corresponding to the infinitesimal generators of the $G$-action. The extended Poisson algebra is given by
\begin{flalign} \label{Extended Poisson}
	&\{f,g \}_{ext} = \hat{i}_{\sigma(\un{V}_f)}\hat{i}_{\sigma(\un{V}_g)}\Omega_{ext} = i_{\un{V}_f}i_{\un{V}_g}\Omega = \{f,g\} \\
	&\{\Phi_{\un{\mu}},\Phi_{\un{\nu}}\}_{ext} = \Phi_{[\un{\mu},\un{\nu}]} \\
	&\{\Phi_{\un{\mu}},f \}_{ext} = \hat{i}_{\Xi_{\un{\mu}}}\hat{i}_{\sigma(\un{V}_f)}\Omega_{ext} = -\hat{i}_{\Xi_{\un{\mu}}}\hat\td f = -\hat{\mathcal{L}}_{\Xi_{\un{\mu}}} f.
\end{flalign}
Again, we stress that the non-extended algebra remains a piece of the extended algebra precisely because the $G$ action was presymplectic and therefore an automorphism of the Poisson bracket. 

The algebra given by \eqref{Extended Poisson} should be interpreted as a ``pre"-crossed product of $\mathcal{M}_{pq}$ and $G$. It consists of the non-extended Poisson algebra $\mathcal{M}_{pq}$, the group algebra\footnote{As would be realized by exponentiating the representation of $\mathfrak{g}$.}, and an action of the group on $\mathcal{M}_{pq}$ via automorphisms of the non-extended Poisson bracket. In these respects, $EPS(\mathcal{M}_{pq},G)$ is algebraically equivalent to the crossed product algebra. The designation ``pre" appears here for the same reason we refer to $\mathcal{M}_{pq}$ as the pre-quantum algebra. In particular, neither $\mathcal{M}_{pq}$ nor $EPS(\mathcal{M}_{pq},G)$ have been realized as von Neumann algebras; i.e., we have not yet constructed a faithful representation of $\mathcal{M}_{pq}$ acting on a suitable Hilbert space. Nevertheless, we have the following very powerful result: the prequantum algebra associated with the extended phase space of $(X,\Omega)$ by the $G$-automorphism $\varphi$ may be regarded as a (pre-) crossed product of the prequantum algebra of $(X,\Omega)$ and $G$. For simplicity, we use the same notation for the pre-crossed product and the crossed product and therefore write:
\beq \label{EPS = CP}
	EPS(\mathcal{M}_{pq},G) = \mathcal{R}(\mathcal{M}_{pq},G).
\eeq

As we will now demonstrate, the pre-crossed product can be promoted to a true crossed product by performing an appropriate quantization of the extended phase space. 

\subsection{Geometric Quantization} \label{sec: GQuant}

{To begin this section, let us describe what we mean when we say an ``appropriate" quantization. Qualitatively, we regard quantization as a functor (structure preserving map) from the category of Poisson algebras into the category of $C^*$ algebras. To formalize this notion, we follow the presentation of \cite{landsman1999lie,landsman2000quantization}, which implements the idea of a so-called \emph{strict} deformation quantization. We begin with a definition:}

\begin{definition}[Field of $C^*$ algebras]
	Let $X$ be a locally compact space, and consider the collection, $\bigg(\mathcal{C},\{A_x,\varphi_x\}_{x \in X}\bigg)$, where $\mathcal{C}$ is a $C^*$ algebra, $A_x$ is a $C^*$ algebra for each $x \in X$, and
	\beq
		\varphi_x: \mathcal{C} \rightarrow A_x
	\eeq
	is a surjective $*$-homomorphism of algebras for each $x \in X$. This collection is called a \emph{continuous field of $C^*$ algebras over $X$} if
	\begin{enumerate}
		\item $x \mapsto \norm{\varphi_x(\mathcal{O})}_{A_x}$ is a continuous map on $X$,
		\item $\norm{\mathcal{O}}_{\mathcal{C}} = \text{sup}_{x \in X} \norm{\varphi_{x}(\mathcal{O})}_{A_x}$, and
		\item for any $f \in C^0(X)$ there exists an element $f\mathcal{O} \in \mathcal{C}$ such that $\varphi_x(f\mathcal{O}) = f(x)\varphi_x(\mathcal{O})$. 
	\end{enumerate}
\end{definition}

Having realized the notion of a field of $C^*$ algebras, we can now state the definition of a \emph{strict deformation quantization}.

\begin{definition}[Strict deformation quantization of a Poisson manifold]
	Let $(P,\{,\})$ be a Poisson manifold, and $I \subseteq \mathbb{R}$ an interval containing $0$ as an accumulation point. A \emph{strict deformation quantization} of $P$ consists of
	\begin{enumerate}
		\item A continuous field of $C^*$ algebras
		\beq
			\bigg(\mathcal{C},\{A_k,\varphi_k\}_{k \in I}\bigg)
		\eeq
		with $A_0 = C^{\infty}(P)$.\footnote{Here $C^{\infty}(P)$ is regarded as an Abelian $C^*$ algebra with composition as the simple product of functions and involution given by complex conjugation.}
		\item A dense subset $\tilde{A_0} \subset A_0$ on which the Poisson bracket is defined -- ergo $\tilde{A_0}$ is a Poisson algebra.
		\item A linear map $Q: \tilde{A_0} \rightarrow \mathcal{C}$ inducing a family of maps $Q_k = \varphi_k \circ Q: \tilde{A_0} \rightarrow A_k$ such that for all $f \in \tilde{A_0}$ and $k \in I$
		\beq
			Q_0(f) = f, \;\; Q_k(f^*) = Q_k(f)^*.
		\eeq
		\item Finally, for all $f,g \in \tilde{A_0}$ Dirac's correspondence condition holds in norm:
		\beq
			\lim_{k \rightarrow 0} \norm{\frac{i}{k}[Q_k(f),Q_k(g)]_{A_k} - Q_k(\{f,g\}_{P})}_{A_k} = 0.
		\eeq
	\end{enumerate}
	If one would like, they can interpret $k \sim \hbar$.   
\end{definition}

{For the purpose of the present work, we focus on a particular quantization scheme that goes under the name of geometric quantization. As originally introduced, geometric quantization is a procedure for constructing a faithful representation of the prequantum algebra of a symplectic manifold $(X,\Omega)$ acting on a bonafide Hilbert space $\mathcal{H}$ \cite{Kostant:1969zz,kostant1970quantization,guillemin1982geometric,
guillemin1984normal,guillemin1990symplectic,Kostant:1987ne}. By extension, we assume that the geometric quantization promotes at least a dense subset of elements of the Poisson algebra of the symplectic manifold into bounded operators on $\mathcal{H}$ thereby realizing a $C^*$ algebra. Given our preference for von Neumann algebras, we can further include within our definition of a geometric quantization a passage from the resulting $C^*$ algebra to its weak closure, thereby realizing a von Neumann algebra associated with the original Poisson algebra.\footnote{{Strictly speaking this introduces new questions about whether the crossed product structure of the Poisson algebra survives the weak closure. For this reason, it may be prudent to omit closure from the quantization scheme with the understanding that the resulting algebra possesses a crossed product structure at the $C^*$ algebra level.}} In practice carrying out a geometric quantization is very difficult; however, for our purpose we will assume that one is possible. In forthcoming work we plan to address the problem of quantization more rigorously rather than taking the existence of a quantization as an assumption.}

We begin with the (pre-) symplectic manifold $(X,\Omega)$. The symplectic form $\Omega$ is termed \emph{integral} if there exists a line bundle $Q \rightarrow X$ with connection $\nabla: TX \rightarrow \text{Der}(Q)$ for which the curvature $R^{\nabla} = \Omega$.\footnote{Alternatively, the integrality of $\Omega$ is equivalent to saying that, as an element of $H^2(X)$ its integrals over homology cycles are integer valued.} An integral symplectic form will also give rise to a Hermitian inner product $g: Q \times Q \rightarrow \mathbb{C}$, which is invariant under parallel transportation by $\nabla$ \cite{kostant1970quantization,guillemin1982geometric}. The triple of data $(Q,\nabla,g)$ is called the \emph{pre-quantum data} on $(X,\Omega)$. 

The prequantum Hilbert space, $\mathcal{H}_{pq}$, is defined to be the space of square integrable sections of $Q$, i.e., $\psi: X \rightarrow Q$. To each Hamiltonian pair $(f,\un{V}_f) \in \mathcal{M}_{pq}$ we can therefore identify an element $\mathcal{O}_f \in \text{Der}(\mathcal{H}_{pq})$ given explicitly by
\beq
	\mathcal{O}_f = -i\nabla_{\un{V}_f} + f.
\eeq	
By construction, $\mathcal{O}: \mathcal{M}_{pq} \rightarrow \text{Der}(\mathcal{H}_{pq})$ is a morphism in the sense that
\beq
	[\mathcal{O}_f, \mathcal{O}_g]_{\text{Der}(\mathcal{H}_{pq})} = i\mathcal{O}_{\{f,g\}}.
\eeq

To complete the geometric quantization scheme we must identify a polarization. In other words, we must identify a Lagrangian subbundle $\mathscr{L} \subset TX$.\footnote{More rigorously, we need a positive definite polarization of the complexified tangent bundle $TX^{\mathbb{C}}$.} Given such a sub-bundle one defines the quantum Hilbert space $(\mathcal{H},g)$ as the set of elements of the prequantum Hilbert space that are covariantly constant along tangent vectors in $\mathscr{L}$:
\beq
	\mathcal{H} = \{\psi \in \mathcal{H}_{pq} \; | \; \nabla_{\un{V}} \psi = 0, \; \forall \; \un{V} \in \mathscr{L}\}.
\eeq
Geometric quantization therefore establishes that our prequantum algebra $\mathcal{M}_{pq}$ may be regarded as a von Neumann algebra
\beq
	GQ(\mathcal{M}_{pq}) \simeq \mathcal{M}_{pq} \subset \mathcal{B}(\mathcal{H}).
\eeq

Notice that the role of geometric quantization is simply to construct the Hilbert space $\mathcal{H}$. 
Geometric quantization therefore commutes with any operation on $\mathcal{M}_{pq}$ that is independent of a chosen representation. As we have established this is true of the crossed product construction. Thus, making use of \eqref{EPS = CP} we establish the following series of equalities:
\beq \label{commutativity achieved}
	GQ \circ EPS(\mathcal{M}_{pq},G) = GQ \circ \mathcal{R}(\mathcal{M}_{pq},G) = \mathcal{R}(GQ(\mathcal{M}_{pq}),G).
\eeq
Equation \eqref{commutativity achieved} is precisely our desired result \eqref{commutative diagram}.

To conclude, let's dig into \eqref{commutative diagram} a little bit more conceptually. In the case that $G$ generates an inner automorphism (or equivalently an equivariant $G$-action on $(X,\Omega)$), the procedure of geometric quantization automatically tells us how to identify the symmetry generators with quantum operators. This is because these generators already belong to $\mathcal{M}_{pq}$ and therefore simply carry over in the geometric quantization of the full prequantum algebra:
\beq
	\Phi_{\un{\mu}} \in \mathcal{M}_{pq} \mapsto \mathcal{O}_{\Phi_{\un{\mu}}} \in GQ(\mathcal{M}_{pq}).
\eeq
In this case constructing the crossed product between $GQ(\mathcal{M}_{pq})$ and $G$ will not change the algebra $GQ(\mathcal{M}_{pq})$ precisely because the symmetry generators already belong to the algebra. Similarly, constructing the extended phase space will not change the prequantum algebra because the $G$-action is already equivariant. Thus, in the case of an inner automorphism we have:
\beq \label{commutativity inner case}
	GQ \circ EPS(\mathcal{M}_{pq},G) = GQ(\mathcal{M}_{pq}) = \mathcal{R}(GQ(\mathcal{M}_{pq}),G)
\eeq  

In the case that $G$ generates an outer automorphism, by contrast, naive geometric quantization of $\mathcal{M}_{pq}$ will not furnish a representation of the symmetry operators since they are not yet contained in $\mathcal{M}_{pq}$. There are two ways that one can proceed to remedy this fact. The first way is to geometrically quantize $\mathcal{M}_{pq}$ to obtain the quantum operator algebra $GQ(\mathcal{M}_{pq})$, carry the presymplectic action of $G$ into an automorphism of $GQ(\mathcal{M}_{pq})$ and then construct the crossed product algebra of $GQ(\mathcal{M}_{pq})$ and $G$ which corresponds to extending $GQ(\mathcal{M}_{pq})$ by an explicit representation of the symmetry generators. The second way is to extend the phase space $(X,\Omega) \mapsto (X_{ext},\Omega_{ext})$ in such a way that the extended $G$-action on $(X_{ext},\Omega_{ext})$ is equivariant. In doing so, we promote the prequantum algebra from $\mathcal{M}_{pq}$ to $EPS(\mathcal{M}_{pq},G)$ and thereby ensure that the symmetry generators belong to $EPS(\mathcal{M}_{pq},G)$ by means of the resulting equivariant moment map. As established in \eqref{EPS = CP}, performing geometric quantization on this extended prequantum algebra realizes the (pre-) crossed product algebra of the non-extended geometrically quantized algebra and $G$. Thus, we again realize our desired result:  
\beq \label{commutativity outer case}
	GQ \circ EPS(\mathcal{M}_{pq},G) = \mathcal{R}(GQ(\mathcal{M}_{pq}),G).
\eeq

\section{Subregions and Entangling Surfaces in Quantum Field Theory} \label{sec: example}

Having laid out the general correspondence between the crossed product and the extended phase space, we are now prepared to offer a physical interpretation of the crossed product which appears in the study of QFT subregions. To this end, we consider the specific implementation of our correspondence to the set of non-extended field configurations on a subregion of spacetime acted on presymplectically by the group of spacetime diffeomorphisms. Our goal is to draw a parallel between the extended phase space by the group of diffeomorphisms, and the crossed product of the subregion algebra with its modular automorphism group. More generally, we will motivate the critical role played by diffeomorphisms of the entangling surface between a subregion and its causal complement. 

For this section we take the following setup: let $M$ be a Lorenzian signature spacetime, and $\Sigma \subset M$ be a Cauchy surface in $M$. Moreover, let $\mathcal{R} \subset \Sigma$ denote an open set inside $\Sigma$ whose closure $\overline{\mathcal{R}}$ is bounded. We will refer to the boundary of $\overline{\mathcal{R}}$ as the corner, $S$, which also plays the role of the entangling surface between $\mathcal{R}$ and its causal complement $\mathcal{R}^c$ \cite{Fliss:2017wop,faulkner2016modular,Fliss:2020cos}. We denote the (pre-) symplectic geometry of non-extended field configurations in the domain of dependence associated with $\mathcal{R}$, $\mathcal{D}(\mathcal{R})$, by $(X_{\mathcal{R}},\Omega_{\mathcal{R}})$, and its associated prequantum algebra of operators by $\mathcal{M}_{pq}$. The geometric quantization of this algebra leads to the non-extended subregion algebra: $GQ(\mathcal{M}_{pq}) \equiv \mathcal{M}_{\mathcal{R}}$.

\subsection{The Diffeomorphism Extended Phase Space} \label{sec: UCS}

As stated, we suppose that $(X_{\mathcal{R}}, \Omega_{\mathcal{R}})$ admits a presymplectic action by $\text{Diff}(M)$. At this point let us pause to address an important distinction. Although we are considering the action of diffeomorphisms on the subregion fields, we are \emph{not} studying a theory of quantum gravity. This is because we have not, and will not, implement the gravitational constraints which gauge the diffeomorphism action.\footnote{With this being said, the considerations we discuss in this note should apply even in quantum gravity, with the primary distinction being that the gravitational constraints must be enforced. We plan to address this problem in future work, with some preliminary steps already being taken in the forthcoming work \cite{CFL2023}.} In this respect, one should regard diffeomorphisms as acting in the background sense. Nonetheless, our central claim is that, even in the non-gravitational context, a subset of diffeomorphisms play a crucial role in completing the specification of physics on a subregion. This is because studying the physics of a subregion requires that one identifies an entanglement cut between the subregion and its complement. However, this entanglement cut cannot be sharply located. As we will demonstrate, the action of diffeomorphisms on the background which have support in the tubular neighborhood of the corner $S$ can be interpreted as `wiggling' the cut around, and have important implications for the physics of the subregion and for the resolution of entanglement singularities. Having made this distinction, let us remark that in this subsection we will make use of intuition from the theory of fully dynamical gravity to inform our analysis in Section \ref{sec: subregionMod} which is valid for generic QFTs.

In an ordinary Lagrangian field theory, the non-extended symplectic geometry generically does not allow for the action of diffeomorphisms to be promoted to an equivariant one \cite{Klinger:2023qna}. 
This is often dismissed based on the observation that diffeomorphisms, in a non-gravitational theory, do not generate physical symmetries because they act on the background. Nevertheless, in the case of a subregion, diffeomorphisms will act non-trivially on the boundary of the subregion. 

It was argued in \cite{Ciambelli:2021vnn} (see also \cite{Freidel:2021cjp,Speranza:2017gxd}) that given a spacetime $M$, there are maximal subalgebras, $\mathcal{A}_k$, of $\text{Diff}(M)$ which are closed under the Lie bracket and that can be associated with the embedding of codimension-$k$ subspaces. In the case $k = 2$, $\mathcal{A}_2$ corresponds to the set of diffeomorphisms which change the embedding of the corner $S$. It has been shown that, for a theory with dynamical gravity, $\mathcal{A}_2$ contains diffeomorphisms that, in general, will support non-zero Noether charges. What's more, all such diffeomorphisms which might support non-zero charges must be contained in $\mathcal{A}_2$. This follows from the well-known result of Noether's second theorem that diffeomorphism charges have support on codimension-2 submanifolds of spacetime. For general codimension-$k$, the aforementioned closed subalgebras take the form 
\beq\label{UCS def}
{\cal A}_k\simeq Lie(\text{Diff}(S_k)\ltimes GL(k,\RR)\ltimes \RR^k).
\eeq
This is the universal corner symmetry (UCS) algebra, with the name inspired by the $k = 2$ case. Notice that the algebra ${\cal A}_2$ consists of the intrinsic diffeomorphisms of $S$, and the generators of transformations (local in $S$) in the group $H\simeq GL(2,\RR)\ltimes \RR^2$. The latter transformations can be thought of as linear endomorphisms and translations in the Lorentzian plane normal to $S$, which will play a significant role in Section \ref{sec: subregionMod}. In view of the special role played by the UCS in theories with dynamical gravity, we will first make remarks relevant to that context, using these as inspiration for considering diffeomorphisms that act in a neighborhood of a subregion boundary in generic quantum field theories.

In past work \cite{Ciambelli:2021nmv, Klinger:2023qna} it has been argued that the proper way to deal with the presence of non-zero, non-integrable charges is to extend the phase space as has been outlined in Section \ref{sec: EPS}. In this case, the extended phase space is equivalent to a principal $\text{Diff}(M)$-bundle, $X^{ext}_{\mathcal{R}}$, over the non-extended configuration space $X_{\mathcal{R}}$. Let $\Phi \in X_{\mathcal{R}}$ correspond to a non-extended field configuration on $\mathcal{D}(\mathcal{R})$, and let $\zeta \in \text{Diff}(M)$ be a diffeomorphism. Viewing $\Phi$ as a map on $M$ with support in $\mathcal{D}(\mathcal{R})$, and $\zeta: M \rightarrow M$, there is a natural action of $\zeta$ on $\Phi$ given by the pullback:
\begin{equation} \label{diff action}
	\varphi_\zeta(\Phi) = \zeta^*\Phi.
\end{equation}
Roughly speaking, we may therefore regard the phase space extended by diffeomorphisms as coinciding with the set of tuples $(\Phi, \zeta)$ where $\Phi$ is a non-extended field configuration, and $\zeta$ is a coordinatization of $M$. 

As we have alluded to, however, most of the degrees of freedom associated with the diffeomorphism $\zeta$ will truly be pure gauge. These degrees of freedom \emph{can} be removed by symplectic reduction. The only degrees of freedom that \emph{are} physical (and therefore survive the symplectic reduction) will be those that correspond to the embedding of the entangling surface $S$. These are the diffeomorphisms that are generated by exponentiating $\mathcal{A}_2$. Thus, as we have addressed in \cite{Klinger:2023qna} any physically relevant diffeomorphism should be regarded as mapping the tubular neighborhood of the embedded corner into $M$, $\zeta: N(S) \rightarrow M$.\footnote{See Appendix \ref{app: Tubular} for a review of the tubular neighborhood expressed as a vector bundle over $S$. Given that ${\cal A}_2$ includes translations (the $\RR^2$ factor in \eqref{UCS def}), this can be viewed as the vector bundle upon which an affine bundle \cite{Ciambelli:2022cfr} is modeled. See further comments in the Conclusion.} In this respect it is more appropriate to view the extended phase space as the set of tuples $(\Phi,\zeta)$ where $\Phi$ is a non-extended field configuration, and $\zeta$ encodes the embedding of the corner $S$ into the bulk spacetime $M$. Alternatively, one may interpret $\zeta$ as a transition map relating the coordinate charts on $\mathcal{R}$ and $\mathcal{R}^c$.

In light of this work, we can now recognize that, upon quantization, the impact of extending the phase space is to cross the non-extended subregion algebra $\mathcal{M}_{\mathcal{R}}$ with the UCS, that is\footnote{Here $\text{UCS}_k = \text{exp}(\mathcal{A}_k)$.}
\beq \label{Extended Operator Algebra}
	GQ \circ EPS(\mathcal{M}_{pq},\text{Diff}(M)) = \mathcal{M}_{\mathcal{R}} \rtimes \text{UCS}_2.
\eeq
Perhaps it seems odd to include the embedding of $S$ among the physical operators that appear in the extended algebra associated with the subregion. However, as we have introduced in Section \ref{sec: extHilb}, these are precisely the degrees of freedom which are required to glue together subregions in a QFT. In this respect, we view the extended phase space as \emph{completing} the subregion algebra by incorporating these crucial degrees of freedom. 

To better appreciate the role played by the UCS on the operator algebra side, we will now carry out a parallel analysis to that which has appeared in this Section, starting from need to consider the crossed product of the non-extended subregion algebra with its modular automorphism group in order to compute entanglement observables in $\mathcal{R}$. In the spirit of the commutative diagram \eqref{commutative diagram}, we will demonstrate how this apparently unrelated analysis leads to the same conclusion as \eqref{Extended Operator Algebra}; the \emph{complete} operator algebra associated with a subregion corresponds to the crossed product of the naive subregion algebra with the UCS.\footnote{We should pause here to note that the UCS, which is a subgroup of the group of diffeomorphisms, is not locally compact (as a group) and thus the construction of the crossed product introduced in Section \ref{sec: CP} is insufficient. In Appendix \ref{app: groupoid} we provide an outline of how this can be remedied by treating the set of diffeomorphisms as a locally compact Lie groupoid for which a left invariant Haar system, and a groupoid $C^*$ algebra can be reliably defined. This allows for the crossed product construction to be generalized into this context. We intend to make use of this fact in forthcoming publications.}

\subsection{The Modular Automorphism Group of a Subregion Algebra} \label{sec: subregionMod}

In Section \ref{sec: UCS}, the issue that predicated the extended phase space was the non-integrability of diffeomorphism charges. In the operator algebra context a similar extension is motivated by the fact that the subregion is described by a type III von Neumann algebra. Thus, as we have addressed in Section \ref{sec: typeIII}, in order to compute finite entanglement observables one must pass from the naive subregion algebra into the crossed product with its modular automorphism group. As we have suggested in Section \ref{sec: extHilb}, there is reason to believe that these two extensions are closely related to each other. In this Section we formalize this connection. In particular, we will study the modular automorphism group of the subregion algebra and the role played therein by shape changing deformations of the entangling surface. As we will uncover, the UCS plays a crucial role in both of these aspects of the operator algebra. Ultimately, we demonstrate how the quantization of the extended phase space \eqref{Extended Operator Algebra} naturally covers the crossed product of the subregion algebra with its modular automorphism group, and allows for state changing deformations to the geometry of the entangling surface. For simplicity, in this section we will work in Minkowski spacetime $M = \mathbb{R}^{1,d-1}$.

Let $\mathcal{H}_{\Sigma}$ denote the Hilbert space associated with the entire Cauchy surface $\Sigma$. For each $\mathcal{R} \subset \Sigma$ there exists a representation $\pi_{\mathcal{R}}: \mathcal{M}_{\mathcal{R}} \rightarrow \mathcal{B}(\mathcal{H}_{\Sigma})$.\footnote{This follows from the isotony assumption of AQFT; given two subregions $\mathcal{R}$ and $\mathcal{R}'$ such that $\mathcal{R} \subset \mathcal{R}'$, it is assumed that $\mathcal{M}_{\mathcal{R}} \subset \mathcal{M}_{\mathcal{R}'}$ \cite{Haag:1963dh}. Thus, the representation $\pi_{\mathcal{R}}$ is equivalent to $\pi_{\Sigma}$ projected down to the subalgebra $\mathcal{M}_{\mathcal{R}}$. In light of this fact we refer to operators simply by $\mathcal{O} \in \mathcal{M}_{\mathcal{R}} \simeq \mathcal{B}(\mathcal{H}_{\Sigma})$, and let the subregion algebra it belongs to be distinguished by the location of the operator insertion.} What's more, under the assumptions of AQFT, the vacuum vector $\ket{\Omega} \in \mathcal{H}_{\Sigma}$ can be shown to be both \emph{cyclic} and \emph{separating} for each $\mathcal{M}_{\mathcal{R}}$. Thus, any state can be realized {as a limit of terms of the form:}
\beq \label{Excited States}
	\ket{\Psi} = \sum_{i} c_{i} \mathcal{O}_{i} \ket{\Omega},
\eeq
for some $c_{i} \in \mathbb{C}$ and $\mathcal{O}_{i} \in \mathcal{M}_{\mathcal{R}}$. In other words, $\ket{\Psi}$ has a path integral preparation with the operators $\mathcal{O}_{i}$ inserted in the subregion $\mathcal{R}$. 

An element $\ket{\Psi} \in \mathcal{H}_{\Sigma}$ defines a weight on $\mathcal{M}_{\mathcal{R}}$ through the expectation value:
\beq
	\omega_{\Psi}(\mathcal{O}) = \bra{\Psi} \mathcal{O} \ket{\Psi}, \; \mathcal{O} \in \mathcal{M}_{\mathcal{R}}.
\eeq
We may therefore construct the modular automorphism group of $\mathcal{M}_{\mathcal{R}}$ associated with the weight $\omega_{\Psi}$.\footnote{See Appendix \ref{sec: mod} for a formal definition of the modular automorphism group of a weight.} The ``reduced density operator" associated with $\omega_{\Psi}$ in the subregion $\mathcal{R}$ is defined by the path integral preparation:
\beq \label{Path Integral Prep of rho}
	\bra{\alpha} \rho_{\mathcal{R}}^{\Psi} \ket{\beta} \propto \int_{\Phi^{-} = \alpha}^{\Phi^{+} = \beta} \text{Vol}_{\mathcal{R}}(\Phi) \sum_{i,j} c^*_i c_j \mathcal{O}_i^{\dagger} \mathcal{O}_j \; e^{-S(\Phi)}.
\eeq
Here $\text{Vol}_{\mathcal{R}}(\Phi)$ is the Liouville measure on $(X_{\mathcal{R}},\Omega_{\mathcal{R}})$, which is diffeomorphism invariant by assumption, and $\ket{\alpha}, \ket{\beta} \in \mathcal{H}_{\Sigma}$ provide the requisite boundary value data. 

In the above we have referred to the ``reduced density operator" in quotes. This is because, strictly speaking, density operators are not well-defined objects in a type III von Neumann algebra which does not admit trace class operators. However, we are only formally defining a reduced density operator as an intermediary step in order to specify the modular Hamiltonian of the weight $\omega_{\Psi}$ which \emph{is} well-defined. By extension, we also refer to the ``subregion modular Hamiltonian", $K^{\Psi}_{\mathcal{R}} = -\ln(\rho^{\Psi}_{\mathcal{R}})$. To define the bonafide modular Hamiltonian on the subregion $\mathcal{R}$, we take the difference of the subregion modular Hamiltonian of $\mathcal{R}$ with that of its causal complement $\mathcal{R}^c$:\footnote{Rigorously, what we have done is to construct the spatial derivative of the weight $\omega_{\Psi}$ by the dual weight specified by the vector $\ket{\Psi}$ acting on the commutant algebra \cite{CONNES1980153}.}
\beq \label{General modular Hamiltonian}
	\hat{K}^{\Psi}_{\mathcal{R}} = K^{\Psi}_{\mathcal{R}} - K^{\Psi}_{\mathcal{R}^c}. 
\eeq

Using the path integral preparation of the modular Hamiltonian implied by \eqref{Path Integral Prep of rho} and \eqref{General modular Hamiltonian}, we will now demonstrate how the modular automorphism group of the vacuum state $\ket{\Omega}$ is generated by an Abelian subalgebra of the UCS for a wide class of subregions. We begin with the most basic subregion which is the half-space:
\beq
	\mathcal{R}_0 = \{x \in \mathbb{R}^{1,d-1} \; | \; x^0 = 0, x^1 > 0\}.
\eeq
Here, we have used the global coordinates $x$ on $\mathbb{R}^{1,d-1}$. The half-space space possesses a non-trivial corner, $S$, located at the intersection between it and its causal complement. We begin with the half space because there exists a theorem, the Bisognano-Wichmann theorem \cite{Bisognano:1976za}, which states that the modular Hamiltonian of the vacuum state associated with any wedge region is the generator of boosts in the Lorentz plane normal to its associated corner, $S$. Thus, we conclude that the modular Hamiltonian of the half-space is indeed an element of the $\mathfrak{so}(1,1)$ subalgebra of $\mathcal{A}_2$. We denote this modular Hamiltonian by $\hat{K}^{\Omega}_{\mathcal{R}_0} \in \mathcal{A}_2$. 

We will now extend the Bisognano-Wichmann theorem to the following: For any subregion $\mathcal{R}_{\zeta}$ which can be realized by deforming the entangling surface of the half-space $\mathcal{R}_0$ by an element $\zeta: N(S) \rightarrow M$ in the UCS, the modular Hamiltonian of the vacuum state is also a UCS generator, $\hat{K}^{\Omega}_{\mathcal{R}_{\zeta}} \in \mathcal{A}_2$. 

To prove this result, we generalize an argument originally made in \cite{faulkner2016modular,faulkner2016shape} that the modular Hamiltonian of the half-space deformed by a diffeomorphism is unitarily equivalent to the modular Hamiltonian of the undeformed half space. Our guiding result is as follows: Let $\mathcal{R}_1$ be a subregion of the Cauchy surface with entangling surface $S$ and modular Hamiltonian $\hat{K}^{\Psi}_{\mathcal{R}_1}: \mathcal{H}_{\Sigma} \rightarrow \mathcal{H}_{\Sigma}$. As indicated, this argument is valid even for the modular Hamiltonian of an excited state, $\ket{\Psi}$ as in \eqref{Excited States}.\footnote{With this being said, for the purpose of proving our desired result it would be sufficient to consider only the modular Hamiltonian of the vacuum.} Suppose that $\zeta: N(S) \rightarrow M$ is a diffeomorphism with support in the tubular neighborhood of the corner, i.e., an element of the UCS. Let $\mathcal{R}_2$ be the image of $\mathcal{R}_1$ under said diffeomorphism. Then, there exists a unitary $U_{\zeta}: \mathcal{H}_{\Sigma} \rightarrow \mathcal{H}_{\Sigma}$ such that the modular Hamiltonian of $\mathcal{R}_2$, $\hat{K}^{\Psi}_{\mathcal{R}_2}: \mathcal{H}_{\Sigma} \rightarrow \mathcal{H}_{\Sigma}$ is unitarily equivalent to $\hat{K}^{\Psi}_{\mathcal{R}_1}$:
\beq \label{Connes cocycle by diffeos}
	\hat{K}^{\Psi}_{\mathcal{R}_1} = U_{\zeta}^{\dagger} \hat{K}^{\Psi}_{\mathcal{R}_2} U_{\zeta}.
\eeq

The proof of \eqref{Connes cocycle by diffeos} is constructive. Let $\{\ket{\Phi}\}$ denote a basis for $\mathcal{H}_{\Sigma}$ corresponding to non-extended field configurations. Then, we can construct the unitary operator \cite{faulkner2016modular}:
\beq \label{unitary}
	U_{\zeta} = \int \text{Vol}_{\mathcal{R}_1}(\Phi_1) \ket{\varphi_{\zeta^{-1}}(\Phi_1)} \bra{\Phi_1}.
\eeq	
Again, $\text{Vol}_{\mathcal{R}_1}$ is the Liouville measure on the symplectic manifold $X_{\mathcal{R}_1}$, and is thus invariant under diffeomorphisms by assumption.  From this fact it is easy to show that \eqref{unitary} is unitary. We have also included a subscript in $\Phi_1$ to remind the reader that $\Phi_1$ should be regarded as a field with domain $\mathcal{R}_1$. The effect of \eqref{unitary} is to replace the field configurations with domain $\mathcal{R}_1$ with field configurations with domain $\mathcal{R}_2$ by directly implementing the action of diffeomorphisms as specified in \eqref{diff action}. In other words the map $\zeta \in \text{Diff}(M) \mapsto U_{\zeta} \in \mathcal{U}(\mathcal{H}_{\Sigma})$ is a unitary representation of the UCS on $\mathcal{H}_{\Sigma}$. 

Let us now demonstrate the validity of \eqref{Connes cocycle by diffeos}. We can easily show that
\beq \label{Unitary on density}
	\rho_{\mathcal{R}_1}^{\Psi} = U_{\zeta}^{\dagger} \rho_{\mathcal{R}_2}^{\Psi} U_{\zeta}.
\eeq
To prove this fact is a simple exercise in path integral manipulation:
\begin{flalign}
	\bra{\alpha_1} \rho_{\mathcal{R}_1}^{\Psi} \ket{\beta_1} &\propto  \int_{\Phi_1^{-} = \alpha_1}^{\Phi_1^{+} = \beta_1} \text{Vol}_{\mathcal{R}_1}(\Phi_1) \sum_{i,j} c^*_i c_j \mathcal{O}_i^{\dagger} \mathcal{O}_j \; e^{-S(\Phi_1)} \\
	&\propto \int_{\varphi_{\zeta}(\Phi^{-}_2) = \alpha_1}^{\varphi_{\zeta}(\Phi^{+}_2) = \beta_1} \text{Vol}_{\mathcal{R}_1}(\varphi_{\zeta}(\Phi_2)) \sum_{i,j} c^*_i c_j \mathcal{O}_i^{\dagger} \mathcal{O}_j \; e^{-S(\varphi_{\zeta}(\Phi_2))} \\
	&\propto \int_{\Phi_2^{-} = \varphi_{\zeta^{-1}}(\alpha_1)}^{\Phi_2^{+} = \varphi_{\zeta^{-1}}(\beta_1)} \text{Vol}_{\mathcal{R}_2}(\Phi_2) \sum_{i,j} c^*_i c_j \mathcal{O}_i^{\dagger} \mathcal{O}_j \; e^{-S(\varphi_{\zeta}(\Phi_2))} \\
	&\propto \bra{\varphi_{\zeta^{-1}}(\alpha_1)} \rho_{\mathcal{R}_2}^{\Psi} \ket{\varphi_{\zeta^{-1}}(\beta_1)} =  \bra{\alpha_1} U_{\zeta}^{\dagger} \rho_{\mathcal{R}_2}^{\Psi} U_{\zeta} \ket{\beta_1}. 
\end{flalign}
Here we have exploited the diffeomorphism invariance of the path integral measure, and we have also used the fact that the diffeomorphism $\zeta$ has support away from the operator insertions $\mathcal{O}_i$ which prepare the state $\ket{\Psi}$.\footnote{In other words, we assume these insertions are deep inside the subregion, away from the entangling surface.} What's more, because we have considered a diffeomorphism with explicit support in the entangling surface between $\mathcal{R}_1$ and its causal complement, the argument \eqref{Unitary on density} is equally valid for the complementary region to $\mathcal{R}_1$. Thus, using \eqref{General modular Hamiltonian}, and the relationship between the subregion modular Hamiltonian and the reduced density operator we conclude \eqref{Connes cocycle by diffeos}, as desired. Equation \eqref{Connes cocycle by diffeos} identifies yet another interpretation of the UCS as the set of diffeomorphisms that implement unitary conjugations of the modular automorphism group of a subregion. This is very closely related to the role played by the UCS in ``gluing" subregions. 

It is now straightforward to conclude our aforementioned result. Let $\mathcal{R}_{\zeta}$ denote the subregion which is obtained from the half-space $\mathcal{R}_0$ by performing a diffeomorphism $\zeta$ which deforms the relevant entangling surface, $S$. Applying \eqref{Connes cocycle by diffeos} to the modular Hamiltonian of the half space in the vacuum state, we find
\beq \label{zeta mod ham}
	\hat{K}^{\Omega}_{\mathcal{R}_{\zeta}} = U^{\dagger}_{\zeta} \hat{K}^{\Omega}_{\mathcal{R}_0} U_{\zeta} = \text{Ad}_{\zeta}(\hat{K}^{\Omega}_{\mathcal{R}_0}),
\eeq
where $\text{Ad}_{\zeta}$ is the adjoint action of $\zeta$ on $\hat{K}^{\Omega}_{\mathcal{R}_0} \in \mathcal{A}_2$. Since $\mathcal{A}_2$ is a closed Lie algebra and the adjoint action is a Lie algebra automorphism, we conclude that $\hat{K}^{\Omega}_{\mathcal{R}_{\zeta}} \in \mathcal{A}_2$. In the most simple case, \eqref{zeta mod ham} allows for the Lorentz plane normal to the corner, and subsequently for the boost associated with the modular Hamiltonian, to be rotated. More generally, \eqref{zeta mod ham} restores the UCS symmetry which is broken by the choice of a particular subtending Cauchy surface.

Equation \eqref{zeta mod ham} puts into focus the relevance of the full UCS in the subregion algebra, beyond simply the fact that the modular Hamiltonian can be realized as one specific infinitesimal generator therein. It makes it clear that the corner symmetry is not only relevant asymptotically (e.g., in the sense that it contains the BMSW symmetries) but it is, in fact, the general symmetry structure underlying entanglement for arbitrary subregions. In future work we plan to demonstrate how other geometrically relevant operators such as the area are naturally included among the charges associated with UCS.

\subsection{A synthesis}

In summary, we have now determined a very broad class of subregions with non-trivial geometry for which the modular Hamiltonian may be regarded as contained in the UCS. We therefore conclude that for such subregions, the crossed product of the non-extended subregion algebra with its modular automorphism group is covered by the crossed product of the non-extended subregion algebra with the UCS. But this is precisely the effect of extending the phase space to accommodate physical diffeomorphisms before quantizing. Thus, we arrive at the conclusion that extending $(X_{\mathcal{R}},\Omega_{\mathcal{R}})$ by $\text{Diff}(M)$ in the presence of an entangling surface -- as one had to do in order to deal with the fact that the $\text{Diff}(M)$ symmetry was non-integrable in $(X_{\mathcal{R}},\Omega_{\mathcal{R}})$ -- prepares the crossed product of $\mathcal{M}_{\mathcal{R}}$ with its modular automorphism group -- which one is forced to construct in order to study finite aspects of the type III von Neumann algebra $\mathcal{M}_{\mathcal{R}}$. In other words, the operator algebra of a bounded subregion is genuinely semi-finite so long as one carefully includes the generators of physical diffeomorphisms from the perspective of the embedded corner associated with the initial value hypersurface.

Still, it may be puzzling why one should take the crossed product with the full UCS rather than just with the modular automorphism group, which is generated by a single element therein. The necessity of the full UCS is really implied by \eqref{Connes cocycle by diffeos}. It is often argued that deforming the shape of a subregion is equivalent to changing the cyclic-separating state which generates the modular automorphism group. As we have shown, such a shape changing deformation results in a conjugation of the modular automorphism group by unitary elements which form a representation of the full UCS. If we would like to interpret this as the effect of a change to the state, that would necessitate that the aforementioned unitary operators are inner relative to the subregion algebra in accordance with the Connes cocycle derivative theorem \eqref{Connes Cocycle Derivative}. However, if that were true in the naive subregion algebra (i.e., before extension) it would also imply that the modular automorphism group of, for example, the half space were inner since it is generated by an element of the same unitary representation of the UCS. Thus, the subregion algebra could not have been type III to begin with. In other words, there is an immediate tension between the desired interpretation, ``shape change = state change", and the general fact that the non-extended subregion algebra is type III. However, this tension is resolved if we take the crossed product of the subregion algebra with the full UCS. Then, not only is the modular automorphism group inner and the algebra is type II, but it is also true that deforming the shape of the subregion can be regarded as a state change generated by the cocycle derivates \eqref{Connes cocycle by diffeos}. Ultimately, if one wishes to have the interpretation of shape changing deformations as implementing a change to the underlying state, the cocycle derivative theorem necessitates that the UCS generators are represented as an inner automorphism regardless of whether the modular automorphism group is generated by a UCS element. This gives a strong motivation for including the full UCS in the crossed product for \emph{any} subregion algebra.

\section{Discussion}

In this note we have provided a new conceptual framework to understand the relationship between the operator algebraic and symplectic geometric approaches to subregions in quantum field theory. Our perspective hinges around an explicit correspondence between the crossed product construction and the extended phase space, which may broadly be interpreted as techniques for implementing symmetries in operator algebras and symplectic geometries, respectively. When combined, these two constructions reveal both an algebraic and a physical mechanism for resolving entanglement singularities which naively appear in subregion operator algebras for QFTs.

From the algebraic perspective, the resolution follows from the dual weight theorem. The dual weight theorem allows for states on a type III algebra to be represented as density operators in the crossed product of that algebra with its modular automorphism group. When applied to subregion algebras, this observation presents a natural approach to assigning entanglement measures to states. To understand the physical mechanism underlying the dual weight theorem, we made use of the correspondence \eqref{commutative diagram}. This correspondence identifies the quantized Poisson algebra associated with the phase space of field configurations extended to accommodate diffeomorphisms with the crossed product algebra \eqref{Extended Operator Algebra}. The algebra \eqref{Extended Operator Algebra} naturally covers the modular automorphism group of the non-extended subregion algebra. Thus, we arrive at the conclusion that the operator algebra associated with a subregion can be rendered semi-finite if one carefully includes the extended degrees of freedom necessitated by the physicality of the UCS subgroup of diffeomorphisms in the presence of corners. Taken together, the dual weight theorem and the quantization of the extended phase space facilitated by the crossed product demonstrate how the inclusion of embedding degrees of freedom associated with entangling surfaces \emph{resolves} the entanglement singularity. This is closely related to the perspective we have motivated in the introduction; rather than eliminating the divergence by placing a cutoff, we have offered a completion to the subregion quantum field theory by restoring degrees of freedom which were missing in the non-extended algebra. Without these degrees of freedom, it is not possible to consistently glue together subregions, resulting in the loss of tensor factorization, and in turn the divergence of the entanglement entropy.

On this note, and as we have alluded to in Section \ref{sec: extHilb}, there is a very close correspondence between the crossed product algebra appearing in the subregion context and the so-called entangling product central to \cite{Donnelly:2016auv}. Recall that the entangling product is of the form
\beq
	\mathcal{H}_{\Sigma}^{ext} = \mathcal{H}_{\mathcal{R}} \otimes_{G_S} \mathcal{H}_{\mathcal{R}^c},
\eeq
where $G_S$ is the group of symmetries which respect the embedding of the entangling surface $S$. We can now naturally identify $G_S = \text{UCS}$. Thus, we propose that the physically motivated crossed product algebra \eqref{Extended Operator Algebra} advocated for in this note should be interpreted as the operator algebraic analog of the extended Hilbert space. Although we have concentrated on the application of our correspondence to quantizing subregion algebras in  theories with a presymplectic diffeomorphism action, the ideas discussed in this note apply equally well to theories in which internal gauge symmetries manifest physically on corners. This is particularly significant in topological field theories, like Chern-Simons theory \cite{Klinger:2023qna}. Quite generically, one may regard the quantization of the extended phase space and its relationship with the crossed product as a manifestation of the fact that gauge theories always possess additional entangling degrees of freedom at the interface between subregions. This is what precludes the tensor factorization of the naive Hilbert space in such theories. In future work, we intend to adapt the crossed product construction and the extended phase space to broader contexts in hopes of better understanding the nature of entanglement in general gauge theories. 

{In regards to the prospect of using the correspondence \eqref{commutative diagram} to understand the quantized operator algebras of gauge theories, we again stress that the analysis presented in the current paper relies on assumptions about the existence of a quantization procedure which lifts a Poisson algebra into a $C^*$ algebra. In future work we plan to address this problem directly by constructing such a quantization from first principles. To this end, the relationship between the Poisson algebra of the extended phase space and the so-called configuration algebroid introduced in Section \ref{sec: EPS} is particularly serendipitous as existing approaches to strict deformation quantization often hinge around viewing quanitzation as derived from the exponentiation of a Lie algebroid to a Lie groupoid \cite{landsman1999lie, landsman2000quantization}. Nevertheless, the correspondence between the extended phase space and the crossed product still presents important lessons about the mathematical structure of gauge theories even in lieu of an explicit quantization map.}

As we were finalizing this note, the works \cite{Jensen:2023yxy,AliAhmad:2023etg} appeared  which also feature the crossed product algebra in the study of subregions. \cite{Jensen:2023yxy} concentrates on subregions in theories with linear graviton fluctuations above a fixed background, extending the approach of \cite{Witten:2021unn}, and provides many clear discussions of related issues including a microscopic computation of the generalized entropy. \cite{AliAhmad:2023etg} recognized the significance of the crossed product in the context of general subregions and worked out several examples. In this note, we have presented a more formal mathematical approach to the crossed product in recognition of its usefulness in analyzing arbitrary type III von Neumann algebras. As we have noted, Takesaki first advocated for using the crossed product of a type III algebra with its modular automorphism group as a representative of the algebra itself in \cite{takesaki1973crossed}. This idea was largely formalized by Haagerup in the construction of his $L^p$ spaces in the series of works \cite{Haagerup1978I,Haagerup1978II,haagerup1979lp}. Using the dual weight theorem, Haagerup developed a non-commutative measure theory for type III von Neumann algebras in which faithful, semi-finite, normal weights are represented as density operators in the crossed product. In light of this, we regard the algebraic contributions of \cite{Jensen:2023yxy,AliAhmad:2023etg} as well as the present note as making the important conceptual leap of applying this mathematical machinery to subregions in QFTs. We would like to highlight that the unique contribution of the present note is that we have not only presented the algebraic significance of the crossed product for subregion QFTs, but we have also emphasized a physical mechanism underlying its significance in the form of the extended phase space and the universal corner symmetry, which are related to the crossed product through the commutative diagram \eqref{commutative diagram}. 

%
The notion of the extended phase space as completing the subregion algebra by supplementing otherwise missing degrees of freedom has some interesting synergy with recent work \cite{Banerjee:2023eew}. In that note, the authors identified the role of Berry phases which appear in the study of Hilbert space representations of operator algebras as providing a topological probe for the type of the algebra. In particular, they argue that Berry holonomies may be interpreted as obstructions to the construction of a trace on the operator algebra in question, and therefore signify the fact that the operator algebra is type III. Applying this observation to the AdS/CFT correspondence, they interpret the type III nature of the naive von Neumann algebra as originating from the existence of different coordinate patches which a local observer does not possess information about. More broadly, they argue that the presence of a geometric phase in the Hilbert space associated with an operator algebra quantifies the fact that the algebra is missing some information. We look forward to more fully understanding the relationship between the present note and this perspective in future work.

\appendix
\renewcommand{\theequation}{\thesection.\arabic{equation}}
\setcounter{equation}{0}

\section*{Acknowledgments}

We thank Luca Ciambelli, Laurent Freidel, Weizhen Jia, Tom Faulkner, and Jonathan Sorce for helpful discussions. A special thank you to Samuel Goldman for providing comments on an earlier version of this note. RGL acknowledges the support of Perimeter Institute for Theoretical Physics where part of this research was carried out. This work was supported by the U.S. Department of Energy under contract DE-SC0015655.

\section{Modular Theory and the GNS Construction} \label{sec: mod}

In this section we review the role of weights in the study of von Neumann algebras, and develop the modular automorphism group of a faithful weight. These ideas play a central role in Section \ref{sec: example}.  

The GNS construction demonstrates how we can associate to any weight $\omega$ a Hilbert space $(\mathcal{H}_{\omega},g_{\omega})$ along with a cyclic vector $\un{v}_{\omega}$ and a faithful representation $\pi_{\omega}: \mathcal{M} \rightarrow \mathcal{B}(\mathcal{H}_{\omega})$. Recall, given a von Neumann algebra $\mathcal{M}$ and a Hilbert space representation $\pi: \mathcal{M} \rightarrow \mathcal{B}(\mathcal{H})$, a vector $\un{\xi}_0 \in \mathcal{H}$ is termed \emph{cyclic} if any element of $\mathcal{H}$ can be realized by acting on $\un{\xi}_0$ with an element of $\pi(\mathcal{M})$. More rigorously, $\pi(\mathcal{M})(\un{\xi}_0)$ is dense in the Hilbert space $\mathcal{H}$. The vector $\un{\xi}_0$ is termed \emph{separating} if $\pi(x)(\un{\xi}_0) = 0$ implies that $x = 0$. 

To build the GNS Hilbert space we first define
\begin{equation}
    N_{\omega} = \{x \in \mathcal{M} \; | \; \omega(x^* x) = 0\},
\end{equation}
which should be interpreted as the kernel of the weight, $\omega$. We may then construct a preclosed inner-product space $\mathcal{H}_{\omega}^0 = \mathcal{M}/N_{\omega}$, which corresponds to the set of equivalence classes of $x \in \mathcal{M}$ up to terms in $N_{\omega}$. Let $\eta_{\omega}(x) \in \mathcal{H}_{\omega}^0$ denote the equivalance class of $x$; we regard $\eta$ as a map from the operator algebra $\mathcal{M}$ into the vector space $\mathcal{H}_{\omega}^0$. Because we have performed the quotient by $N_{\omega}$ we can construct a non-degenerate inner product on $\mathcal{H}_{\omega}^0$ through the weight $\omega$ as
\begin{equation} \label{GNS inner}
    g_{\omega}(\eta_{\omega}(x),\eta_{\omega}(y)) = \omega(y^* x).
\end{equation}
To upgrade $\mathcal{H}_{\omega}^0$ into a Hilbert space, we have to ensure that it is closed in the topology generated by \eqref{GNS inner}. Thus, we define the GNS Hilbert space of $\omega$, $\mathcal{H}_{\omega}$, to be the closure of $\mathcal{H}_{\omega}^0$ under this inner product. 

Given the GNS Hilbert space $\mathcal{H}_{\omega}$ we define the vector $\un{v}_{\omega} = \eta_{\omega}(\mathbb{1}) \in \mathcal{H}_\omega$, which satisfies the condition that:
\begin{equation}
    g_{\omega}(\eta_{\omega}(x),\un{v}_{\omega}) = \omega(x), \forall x \in \mathcal{M}.
\end{equation}
We also define the representation $\pi_{\omega}: \mathcal{M} \rightarrow \mathcal{B}(\mathcal{H}_{\omega})$ using the composition law native to $\mathcal{M}$:
\begin{equation} \label{GNS Rep}
    \pi_{\omega}(x)(\eta_{\omega}(y)) = \eta_{\omega}(xy).
\end{equation}
Equation \eqref{GNS Rep} implies that $\pi_{\omega}(x)(\un{v}_{\omega}) = \eta_{\omega}(x)$, which allows us to conclude that $\un{v}_{\omega}$ is, indeed, a cyclic vector. Notice, also, that it allows us to realize a more down to earth interpretation of the weight:
\begin{equation}
    \omega(x) = g_{\omega}(v_{\omega}, \pi_{\omega}(x)(v_{\omega})) \equiv \bra{v_{\omega}} \pi_{\omega}(x) \ket{v_{\omega}}.
\end{equation}
Here we have used the bra-ket notation familiar from quantum mechanics to stress that, in the GNS Hilbert space associated with the weight $\omega$, the computation of the weight applied to an operator $x \in \mathcal{M}$ corresponds to the expectation value of the operator $x \in \mathcal{M}$ evaluated in the state $\ket{v_{\omega}} \in \mathcal{H}_{\omega}$.

If $\omega$ is faithful we can moreover show that $\un{v}_{\omega}$ is separating. This follows more or less by definition. For a generic weight, $\pi_{\omega}(x)(\un{v}_{\omega}) = 0$ implies that $\eta_{\omega}(x) = 0$. In other words $x$ is an element of the kernel $N_{\omega}$, but may not be explicitly equal to zero. However, if $\omega$ is faithful, $\omega(x^*x) = 0$ directly implies that $x = 0$. Thus, in this case we have separability in addition to cyclicity.

Let $\varphi$ be a faithful weight, and let $(\pi_{\varphi}, \mathcal{H}_{\varphi}, \eta_{\varphi})$ denote its GNS construction. Let $\un{v}_{\varphi}$ denote the cylic-separating vector in $\mathcal{H}_{\varphi}$ associated with $\varphi$. We define the \emph{Tomita operator}
\beq
	S_{\varphi}: \mathcal{H}_{\varphi} \rightarrow \mathcal{H}_{\varphi}; \; \eta_{\varphi}(x) \mapsto \eta_{\varphi}(x^*). 
\eeq
In other words, $S_{\varphi}$ carries the involution on $\mathcal{M}$ into the Hilbert space $\mathcal{H}_{\varphi}$. The Tomita operator possesses a unique polar decomposition as
\beq
	S_{\varphi} = J_{\varphi} \Delta_{\varphi}^{1/2},
\eeq
where here $\Delta_{\varphi}$ is a linear, positive, non-singular, self adjoint operator on $\mathcal{H}_{\varphi}$ called the \emph{Modular Operator}, and $J_{\varphi}$ is an antilinear isometry on $\mathcal{H}_{\varphi}$ called the \emph{Modular Conjugation Operator}. Tomita's seminal result establishes that the modular operator defines a one parameter family of automorphisms on $\mathcal{M}$, $\sigma^{\varphi}: \mathbb{R} \times \mathcal{M} \rightarrow \mathcal{M}$:
\beq
	\pi_{\varphi} \circ \sigma_t^{\varphi}(x) = \Delta_{\varphi}^{it} \pi_{\varphi}(x) \Delta_{\varphi}^{-it}.
\eeq
This is the modular automorphism group \cite{Takesaki:1970aki,Combes1971}. 

The weight $\varphi$ and the automorphism group $\sigma_t^{\varphi}$ satisfy the modular compatibility conditions:
\begin{enumerate}
    \item The weight $\varphi$ is invariant under $\sigma^{\varphi}$: $\varphi = \varphi \circ \sigma^{\varphi}_{t}$ for all $t \in \mathbb{R}$.
    \item For every pair $x,y \in \mathcal{M}$ there exists a bounded function, $F_{x,y}: \overline{D} \rightarrow \mathbb{C}$, whose domain is the closed horizontal strip $\overline{D} \subset \mathbb{C}$ bounded by $\mathbb{R}$ and $\mathbb{R} + i$, and which is holomorphic on the open strip $D$ satisfying:
    \begin{equation}
        F_{x,y}(t) = \varphi(\sigma^{\varphi}_t(x)y), \qquad F_{x,y}(t+i) = \varphi(y \sigma^{\varphi}_t(x)).
    \end{equation}
\end{enumerate}
The second of these two conditions is called the Kubo-Martin-Schwinger (KMS) condition \cite{kubo1957,Martin:1959jp,Haag:1967sg}. In quantum statistical mechanics, it characterizes when a weight corresponds to the thermal equilibirum state of a quantum statistical ensemble at inverse temperature $\beta = -1$. Thus, we arrive at the physical interpretation of the modular flow as a kind of generalized thermal evolution operator, and of the associated weight as the equilibrium state associated with this evolution. Associated with this interpretation is the identification of the infinitesimal generator of the modular automorphism group
\beq
	\Delta_{\varphi} = \text{exp}(-K^{\varphi}).
\eeq
Here $K^{\varphi}$ is  the \emph{Modular Hamiltonian} associated with the weight $\varphi$.

To close, let us discuss the relationship between different faithful weights on a common von Neumann algebra. Indeed, one aspect of the preceding analysis that may have given the reader some pause is the reliance on a chosen faithful weight, $\varphi$, in order to develop the non-commutative measure theory of a von Neumann algebra $\mathcal{M}$. From a physical perspective this seems like fixing a single state. However, we will now state a powerful set of results typically attributed to Connes \cite{connes1973theoreme} which ensure that the GNS construction of any single faithful weight contains all of the information in every faithful weight on $\mathcal{M}$. 

Again, let $\varphi$ be a faithful, semifnite, normal weight on the von Neumann algebra $\mathcal{M}$, and let $(\pi_{\varphi}, \mathcal{H}_{\varphi},\eta_{\varphi})$ denote its GNS Hilbert representation. For any other faithful, semi-finite, normal weight $\psi$ on $\mathcal{M}$ with GNS representation $(\pi_{\psi}, \mathcal{H}_{\psi}, \eta_{\psi})$, there exists an isometry $U: \mathcal{H}_{\psi} \rightarrow \mathcal{H}_{\varphi}$ which intertwines the GNS representations of $\varphi$ and $\psi$:
\beq
	\pi_{\varphi}(x)\circ U = U \circ \pi_{\psi}(x), \; \forall  x \in \mathcal{M}.
\eeq
A corollary of this fact is that for any faithful, semi-finite, normal weight $\psi \in \mathcal{M}_*$ there exists a vector $\xi_{\psi} \in \mathcal{H}_{\varphi}$ such that
\begin{equation}
    \psi(x) = g_{\varphi}(\xi_{\psi},\pi_{\varphi}(x)(\xi_{\psi})) \equiv \bra{\xi_{\psi}} \pi_{\varphi}(x) \ket{\xi_{\psi}}.
\end{equation}
In other words, each faithful weight, $\psi$, is associated uniquely with a state $\ket{\xi_{\psi}} \in \mathcal{H}_{\varphi}$. Finally, we have the problem of relating the modular automorphism groups of two faithful weights. This relationship is formalized by the Connes' cocycle derivative theorem. Given two faithful weights $\varphi$ and $\psi$, there exists a unique continuous one parameter family of unitary elements $\{u_t\} \subset \mathcal{M}$ such that
\beq \label{Connes Cocycle Derivative}
	\sigma_t^{\varphi}(x) = u_t \sigma^{\psi}_{t}(x) u_t^*, \; \forall t \in \mathbb{R}, x \in \mathcal{M}.
\eeq
Thus, the modular automorphism groups of any two faithful weights are \emph{inner} unitarily equivalent.

\section{Canonical Purification and the general structure of a von Neumann algebra} \label{sec: canon}

The content of this section may be regarded as a simplified account of the Haagerup's $L^p$ spaces \cite{haagerup1979lp}.\footnote{Haagerup's construction makes use of the dual action relative to the modular automorphism group, and the resulting scaling property of the trace defined on the crossed product algebra. We omit these details for the sake of brevity.}

Let $\mathcal{M}$ be a von Neumann algebra of arbitrary type, and let $\mathcal{N}_{\mathcal{M}} = \mathcal{M} \rtimes_{\sigma} \mathbb{R}$ denote the crossed product of $\mathcal{M}$ by its modular automorphism group. In the case that $\mathcal{M}$ is semi-finite, there is an isomorphism between $\mathcal{M}$ and $\mathcal{N}_{\mathcal{M}}$, and one may simply regard the following analysis as taking place in $\mathcal{M}$. In the case that $\mathcal{M}$ is a type III von Neumann algebra, however, $\mathcal{N}_{\mathcal{M}}$ is its type II representative. 

Because $\mathcal{N}_{\mathcal{M}}$ is a semi-finite algebra, it possesses a semi-finite, normal, faithful trace $\tau: \mathcal{N}_{\mathcal{M}} \rightarrow \mathbb{C}$. A density operator $\rho \in \mathcal{N}_{\mathcal{M}}$ is then an element of $\mathcal{N}_{\mathcal{M}}$ which is positive, self-adjoint, and normalized i.e. $\tau(\rho) = 1$. We will suppose that there exists a faithful representation $\pi: \mathcal{N}_{\mathcal{M}} \rightarrow \mathcal{B}(\mathcal{H})$ for some Hilbert space $\mathcal{H}$, and for which the trace $\tau$ can be related to the Hilbert space trace $tr_{\mathcal{H}}$ as
\beq
	\tau(x) = tr_{\mathcal{H}} \circ \pi(x). 
\eeq

Let $(\pi_{\tau}, \mathcal{H}_{\tau}, \eta_{\eta})$ denote the GNS Hilbert space associated with the trace $\tau$. This Hilbert space is naturally identified with the Hilbert space of operators of Hilbert-Schmidt class, in other words $\mathcal{H}_{\tau} \simeq \mathcal{H} \otimes \mathcal{H}^* \equiv \mathcal{H}_{HS}$. Indeed, by construction the inner product on $\mathcal{H}_{\tau}$ is equivalent to the Hilbert-Schmidt inner product on $\mathcal{H}_{HS}$:
\beq
	g_{\tau}(\eta_{\tau}(x), \eta_{\tau}(y)) = \tau(y^*x) = tr_{\mathcal{H}}(\pi(y)^{\dagger} \pi(x)). 
\eeq
Given $x \in \mathcal{N}_{\mathcal{M}}$ we may therefore write:
\beq
	\tau(x) = tr_{\mathcal{H}}(\pi(x)) = g_{\tau}(\un{v}_{\tau}, \pi_{\tau}(x)(\un{v}_{\tau})),
\eeq
where $\un{v}_{\tau} \in \mathcal{H}_{\tau}$ is the cyclic-separating vector associated with $\tau$. 

Every density operator in $\mathcal{N}_{\mathcal{M}}$ may be represented by a unique element $\un{\psi}_{\rho} \in \mathcal{H}_\tau$. This follows from the fact that density operators are positive and self adjoint. Firstly, by positivity, there exists an operator $\rho^{1/2} \in \mathcal{N}_{\mathcal{M}}$ such that $\rho = \rho^{1/2}\rho^{1/2}$. Let us define:
\beq
	\un{\psi}_{\rho} = \pi_{\tau}(\rho)^{1/2}(\un{v}_{\tau}) = \eta_{\tau}(\rho^{1/2}).
\eeq
Then, using the fact that $\rho$ is a self adjoint operator, we can write:\footnote{Here we have used the cylicity of the trace.}
\beq
	\tau(\rho x) = tr_{\mathcal{H}}(\pi(\rho) \pi(x)) = tr_{\mathcal{H}}(\pi(\rho^{1/2}) \pi(x) \pi(\rho^{1/2})) = g_{\tau}(\un{\psi}_{\rho}, \pi_{\tau}(x)(\un{\psi}_{\rho})).
\eeq	
In the physics literature, this procedure is referred to as the canonical purification of the density operators in $\mathcal{N}_{\mathcal{M}}$. In particular, each ``mixed state" $\rho$ has been given representation as a ``pure state" $\un{\psi}_{\rho}$.

Because $\mathcal{N}_{\mathcal{M}}$ is semi-finite, one can moreover show that each faithful, normal, positive weight, $\varphi$, on $\mathcal{N}_{\mathcal{M}}$ may be uniquely represented by a density operator $\rho_{\varphi} \in \mathcal{N}_{\mathcal{M}}$ such that
\beq
	\varphi(x) = \tau(\rho_{\varphi} x) = tr_{\mathcal{H}}(\pi(\rho_{\varphi})\pi(x)), \; \forall x \in \mathcal{N}_{\mathcal{M}}.
\eeq
Thus, by the preceding discussion, such a weight may also be represented by a vector $\un{\psi}_{\varphi} \equiv \un{\psi}_{\rho_{\varphi}} \in \mathcal{H}_{\tau}$, for which
\beq
	\varphi(x) = g_{\tau}(\un{\psi}_{\varphi}, \pi_{\tau}(x)(\un{\psi}_{\varphi})). 
\eeq
Since $\varphi$ is a faithful weight, this vector can be shown to be both cyclic and separating for $\mathcal{N}_{\mathcal{M}}$. 

In the case that $\mathcal{M}$ was a type III von Neumann algebra to begin with, we can now make use of the dual weight theorem in order to identify a weight $\varphi$ on $\mathcal{M}$ with a density operator in $\mathcal{N}_{\mathcal{M}}$. In particular, let $\tilde{\varphi}$ denote the weight on $\mathcal{N}_{\mathcal{M}}$ which is dual to $\varphi$ a weight on $\mathcal{M}$. Then, $\varphi$ is identified with the density operator $\rho_{\tilde{\varphi}} \in \mathcal{N}_{\mathcal{M}}$. The entanglement entropy of this density operator may be regarded as the generalized entropy associated with the weight $\varphi$. 

Finally, in this context we can provide very explicit forms for the Tomita operator, the modular operator, and the modular Hamiltonian associated with a faithful weight. To begin, the Tomita operator is given by
\beq \label{Tomita in canonical purification}
	S_{\varphi}(x) = \rho_{\varphi}^{-1/2} x^* \rho_{\varphi}^{1/2}.
\eeq
From \eqref{Tomita in canonical purification}, we can see that the modular operator in the canonical purification picture is identifiable with the density operator associated to the weight: $\Delta_{\varphi} = \rho_{\varphi}$. The modular automorphism group is therefore given by
\beq \label{Modular Aut in canonical purification}
	\sigma^{\varphi}_t(x) = \rho_{\varphi}^{it} x \rho_{\varphi}^{-it},
\eeq
which we naturally interpret as Heisenberg evolution with respect to the density operator $\rho_{\varphi}$. The modular Hamiltonian is the infinitesimal generator of this evolution:
\beq \label{Modular Ham in canonical purification}
	K^{\varphi} = -\ln(\rho_{\varphi}). 
\eeq

\section{Generalized Conditional Expectations} \label{app: condexp}

In this section we briefly introduce the generalized conditional expectation of Accardi and Cecchini. We also construct the conditional expectation dual to the embedding of a von Neumann subalgebra.

To begin, let $M$ and $N$ be two von Neumann algebras, with $\gamma: N \rightarrow M$ a unital, positive, linear map between them. Let $\omega_M$ be a faithful weight on $M$, and define $\omega_N = \omega_M \circ \gamma$, which is a weight on $N$. In what follows, $(\pi_i, \mathcal{H}_i, \eta_i)$ denotes the GNS Hilbert spaces of $\omega_i$ with $i = M$ or $N$, and $\un{\xi}_i \in \mathcal{H}_i$ is vector representing $\omega_i$. 

The following theorem is due to Accardi and Cecchini, and constitutes a definition of the generalized conditional expectation \cite{accardi1982conditional}. Provided $\omega_N$ is itself a normal and faithful weight on $N$, there exists a unique unital, normal, and faithful map $\gamma': M' \rightarrow N'$ such that 
\begin{equation} \label{AC theorem}
    g_M(\pi_M \circ \gamma(n)(\un{\xi}_M), \mu(\un{\xi}_M)) = g_N(\pi_N(n)(\un{\xi}_N),\gamma'(\mu)(\un{\xi}_N)), \qquad \forall n \in N, \mu \in M'. 
\end{equation}
In other words, $\gamma'$ is a generalized adjoint operator that passes to a map on the commutant algebras. Let $J_i$ denote the modular conjugation operators associated with the pair of weights. Then, we can define the operator
\begin{equation}
    \gamma^*: M \rightarrow N, \; \gamma^*(m) = J_N \gamma'(J_M \pi_{M}(m) J_M) J_N. 
\end{equation}
which enjoys the following properties:
\begin{enumerate}
    \item $\omega_N \circ \gamma^* = \omega_M$, 
    \item $g_M(J_M \pi_M(m)(\un{\xi}_M),\pi_M \circ \gamma(n)(\un{\xi}_M)) = g_N(J_N \pi_N \circ \gamma^*(m)(\un{\xi}_N),\pi_N(n)(\un{\xi}_N))$ for each $m \in M$ and $n \in N$.
    \item $\gamma^*(\mathbb{1}_{M}) = \mathbb{1}_N$
\end{enumerate}
The map $\gamma^*$ is called the $\omega_M$-dual map of $\gamma$, and defines the generalized conditional expectation.

Notice that the necessary and sufficient condition to proceed from the map $\gamma$ to the generalized conditional expectation $\gamma^*$ is that the weight $\omega_N = \omega_M \circ \gamma$ is normal and faithful. We will now demonstrate that this condition is always met when $\gamma$ corresponds to the embedding of a von Neumann subalgebra. We also construct the resulting generalized conditional expectation. 

Suppose $N \subset M$ is a von Neumann subalgebra, and let $\gamma: N \hookrightarrow M$ denote an associated injection. Suppose that $M$ is represented in a Hilbert space $\mathcal{H}$ with a cyclic-separating vector $\un{\Omega}$ associated with the faithful weight:
\begin{equation}
    \omega_{\Omega}(x) = g(\un{\Omega}, \pi(x)(\un{\Omega})), \; x \in M. 
\end{equation}
Let 
\begin{equation}
    P_N: \mathcal{H} \rightarrow \mathcal{H}_{N}
\end{equation}
denote the orthogonal projection induced by $\gamma$. Here $\mathcal{H}_N \equiv \overline{\pi(N)(\un{\Omega})}$ is the Hilbert space obtained by acting on $\un{\Omega}$ with elements of the subalgebra. Thus, $\un{\Omega}$, treated as an element of $\mathcal{H}_N$, is cyclic and separating by construction. In turn, the weight
\begin{equation}
    \omega_{\Omega}\rvert_N(n) = g(\un{\Omega}, \pi(n)(\un{\Omega})), \; n \in N
\end{equation}
remains normal and faithful for $N$, and thus the existence of a generalized conditional expectation dual to $\gamma$ is guaranteed. 

Let $J_M: \mathcal{H} \rightarrow \mathcal{H}$ and $J_N: \mathcal{H}_N \rightarrow \mathcal{H}_N$ denote the modular conjugation operators associated with $\omega_{\Omega}$ and $\omega_{\Omega}\rvert_N$, respectively. Then, we have the following explicit form for the $\omega_{\Omega}$-dual map of the injection $\gamma$:
\begin{equation}
    \gamma^*_{\Omega}: M \rightarrow N, \; \pi \circ \gamma^*_{\Omega}(m) = J_N P_N J_M \pi(m) J_M P_N J_N. 
\end{equation}
This map satisfies:
\begin{equation}
    \omega_{\Omega}\rvert_N \circ \gamma^*_{\Omega}(m) = \omega_{\Omega}(m),
\end{equation}
and defines a generalized conditional expectation from $M$ onto $N$. 

\section{The Tubular Neighborhood Theorem} \label{app: Tubular}

Let $S_k \subset X$ denote a codimension-$k$ submanifold of a smooth $d$-dimensional manifold $X$. A tubular neighborhood of $S_k$ in $X$ is a vector bundle $\pi: N(S_k) \rightarrow S_k$ together with a smooth map $J: N(S_k) \rightarrow X$ satisfying the following properties:
\begin{enumerate} 
	\item Let $z_{N(S_k)}: S_k \rightarrow N(S_k)$ denote the zero section of $N(S_k)$. Then,
	\beq \label{Tubular Neighborhood}
		J \circ z_{N(S_k)}: S_k \rightarrow X
	\eeq
	is an embedding of $S_k$ into $M$.
	\item There exist open subsets $U \subseteq N(S_k)$ and $V \subseteq X$ such that $\text{im}(z_{N(S_k)}) \subset U$ and $S_k \subset V$, and for which $J\rvert_{U}$ is a diffeomorphism. 
\end{enumerate}   
The tubular neighborhood theorem ensures the existence of such a tubular neighborhood for any embedded submanifold of $X$ \cite{bott1982differential}.

One should interpret $J$ as corresponding to a coordinate chart for the tubular neighborhood. Schematically, we regard the local coordinate system as given by $(\sigma^{\alpha}, u^a)$ with $\alpha = 1, ..., d-k$ and $a = 1, ..., k$. Here $\sigma^{\alpha}$ is a coordinate system native to the manifold $\spac{k}$, while $u^a$ are fiber coordinates parameterizing the directions normal $S_k$ within the tubular neighborhood. The tubular neighborhood is independent of the $d$-dimensional bulk into which it is embedded. It should be thought of as the $d$-dimensional manifold which can be canonically constructed starting from some $(d-k)$-dimensional manifold $S_k$.

\section{Cross Products with Locally Compact Groupoids} \label{app: groupoid}

{In order to implement the crossed product for infinite dimensional groups, like the group of diffeomorphisms and its UCS subgroups, we must pass to the context of a Lie groupoid.  The set of diffeomorphisms is not locally compact as a group (and thus a suitable Haar measure does not exist), but the pair groupoid --- which coincides with the set of diffeomorphisms --- is locally compact when viewed as a topological groupoid.\footnote{In fact, every Lie groupoid is locally compact as a topological groupoid, see \cite{paterson2012groupoids} for a useful discussion.} For any Lie groupoid we can construct a left Haar system, which is a collection of Lebesgue measures generalizing the left Haar measure of a Lie group. By extension, every Lie groupoid gives rise to a groupoid $C^*$ algebra which is sufficient for generalizing the crossed product. We expect that this remark will be particularly important if one wishes to view gravitational theories through the same lens as other gauge theories. 

In this section we provide a brief introduction to the aspects of Lie groupoids which are necessary to generalize the notion of the crossed product to group structures like the set of diffeomorphisms. The reader is referred to \cite{landsman1999lie,landsman2000quantization,paterson2012groupoids} for a more complete introduction.} 

\begin{definition}[Groupoid]
	A \emph{groupoid} consists of a pair of sets, $G_1$ and $G_0$, along with a collection of maps
	\begin{flalign} \label{Groupoid maps}
		&s,t: G_1 \rightarrow G_0 \; \text{(Source and Target Projections)}, \\
		&m: G_1^{(2)} \rightarrow G_1 \; \text{(Multiplication)}, \\
		&u: G_0 \rightarrow G_1 \; \text{(Unit Element)}, \\
		&i: G_1 \rightarrow G_1 \; \text{(Inversion)}.
	\end{flalign}
The set $G_0$ is called the \emph{object set}, while the set $G_1$ is called the \emph{morphism set}. The groupoid itself may be regarded as a product structure $G \simeq G_0 \times G_1 \times G_0$ such that an element $\varphi \in G$ is of the form $\varphi = (t(g),g,s(g))$. In this respect one should regard $g \in G_1$ as a map, or arrow, from its source $s(g)$ into its target $t(g)$. The set
\beq \label{composable arrows}
	G_1^{(2)} \equiv \{(g,h) \in G_1 \; | \; s(g) = t(h)\},
\eeq
first appearing in \eqref{Groupoid maps} may be regarded as defining the set of composable arrows. Composition in the groupoid is specified by the map $m$ which we denote by
\beq
	m((g,h)) = gh, \; s(gh) = s(h), \; t(gh) = t(g). 
\eeq
The composition law must be associative wherever multiple composition of arrows makes sense:
\beq
	g(hl) = (gh)l, \; \forall (g,h),(h,l),(g,hl),(gh,l) \in G_1^{(2)}.
\eeq
The map $u: G_0 \rightarrow G_1$ embeds the object set into the morphism manifold and acts as a unit element when composed with arrows in an appropriate manner. That is
\beq
	s(u(x)) = t(u(x)) = x, \; gu(s(g)) = g, \; u(t(g)) g = g. 
\eeq
Finally, the map $i: G_1 \rightarrow G_1$ defines an inverse operation which we denote by
\beq
	i(g) = g^{-1}, \; s(g^{-1}) = t(g), \; t(g^{-1}) = s(g).
\eeq
The inversion satisfies the expected property:
\beq
	g^{-1}g = u(s(g)), \; gg^{-1} = u(t(g)). 
\eeq
\end{definition}

A groupoid is called a \emph{Lie groupoid} if its sets of objects $G_0$ and morphisms $G_1$ are both manifolds, and the maps $s,t,m,u,i$ are smooth, and the maps $s,t$ are submersions. Given a Lie groupoid, we define
\beq
	G_1^{x} \equiv s^{-1}(x) \cap t^{-1}(x) = \{g \in G_1 \; | \; t(g) = s(g) = x\}.
\eeq
It follows immediately from the definition of the Lie groupoid that $G_{1}^{x}$ has the structure of a Lie group at each $x \in G_0$. These are called the \emph{isotropy groups} of the groupoid $G$. 

Groupoids are rather complicated objects, so we give a few simple examples to orient ourselves. In what follows we will often refer to a groupoid as $G = G_1 \rightrightarrows G_0$ or simply $G_1 \rightrightarrows G_0$. 

\begin{example}[Lie group and Bundle of Lie groups]
	A Lie group is an example of a Lie groupoid with a trivial manifold of objects. In other words $G_1 = H$ is a Lie group and $G_0 = \{x\}$. Similarly, one can imagine a case in which $t = s$ in which case $G_1$ is exactly equal to the set of all its isotropy groups. In this case $G_1 \rightarrow G_0$ may be regarded as a bundle of Lie groups over $G_0$. 
\end{example}

\begin{example}[Pair groupoid]
	The pair groupoid has $G_1 = M \times M$ and $G_0 = M$ where $M$ is a manifold. Each $g \in G_1$ is merely regarded as a trivial arrow between two specified points in $M$: $(x,y) \in G_1$ with $t(x,y) = x$ and $s(x,y) = y$. Composition, unit, and inversion are also quite easily determined:
	\begin{flalign}
		&(x,y)(y,z) = (x,z), \\
		&u(x) = (x,x), \\
		&(x,y)^{-1} = (y,x). 
	\end{flalign}
\end{example}

\begin{example}[Action groupoid]
	The action groupoid is in some sense a minimal generalization of the pair groupoid. Let $M$ be a manifold and let $\varphi: H \times M \rightarrow M$ be a Lie group action on $M$. Then, $(g,x) \in G_1 \equiv H \times M$ may be regarded as an element of the Lie groupoid with $s(g,x) = x$ and $t(g,x) = \varphi_g(x)$. 
\end{example}

\begin{example}[General linear groupoid]
	Let $E \rightarrow M$ be a vector bundle. Then $G_1 = Gl(E)$ and $G_0 = M$ define a groupoid whose elements are linear endomorphisms between the fibers of the bundle $E$. That is,
	\beq
		Gl(E) \ni g: E_{s(g)} \rightarrow E_{t(g)}.
	\eeq
\end{example}

\begin{example}[Gauge groupoid]
	Our final example is a particularly important one -- the \emph{gauge groupoid}. Given a principal bundle $P \rightarrow M$ with structure group $H$, we can define an associated Lie groupoid with $G_1 = (P \times P)/G$ and $G_0 = M$. 
\end{example}

The gauge groupoid is important for the following reason,
\begin{definition}[Transitivity]
	A Lie groupoid is called \emph{transitive} if the map $(s,t): G_1 \rightarrow G_0 \times G_0$ is surjective. Any transitive Lie groupoid is isomorphic to a gauge groupoid for some choice of principal bundle. In particular, this is the principal bundle $s^{-1}(x) \rightarrow G_0$ with structure group $G_1^{x}$ for any choice of $x \in G_0$. 
\end{definition}

This means that we can understand any transitive groupoid by passing to a related principal bundle, and visa-versa. For example, the general linear groupoid is transitive and can be realized as the gauge groupoid associated with the principal frame bundle of the vector bundle $E$. 

\begin{definition}[Lie groupoid representations]
	Given a Lie groupoid $G_1 \rightrightarrows G_0$ and a vector bundle $E \rightarrow G_0$ we can define a Lie groupoid representation
	\beq
		\rho: G_1 \rightarrow Gl(E), \; s_{Gl}\circ \rho(g) = E_{s(g)}, \; t_{Gl} \circ \rho(g) = E_{t(g)}. 
	\eeq
\end{definition}

A salient feature of a (locally compact) Lie group is that it comes equipped with a canonical measure. This, too, can be generalized to Lie groupoids. 

\begin{definition}[Left invariant vector fields]
	Let $G = G_1 \rightrightarrows G_0$ be a Lie groupoid. In the case of a Lie group, the Lie algebra is defined by first identifying the set of left invariant vector fields. We will follow the same course here. To begin, we define the left action:
	\beq \label{Groupoid left action}
		L_g: G_1^{s(g)} \rightarrow G_1^{t(g)}, \; L_g(g') = gg'.
	\eeq
	Notice that \eqref{Groupoid left action} maps elements in the isotropy bundle of the source to elements of the isotropy bundle of the target. Since $G_1$ is a manifold, there is nothing ambiguous about defining its tangent bundle $TG_1$. Along the identity section, the fibers of this tangent bundle decompose as
	\beq
		T_{u(x)} G_1 = T_x G_0 \oplus \text{ker}(t_*) \rvert_{u(x)}.
	\eeq
Given a tangent vector $\un{X} \in TG_1$ we say that it is \emph{left invariant} if
	\beq
		t_* \un{X} = 0, \text{ and } (L_g)_* \un{X}\rvert_{g'} = \un{X}\rvert_{gg'}. 
	\eeq
\end{definition}

\begin{definition}[Left Haar system]
	A \emph{left Haar system} on a Lie groupoid $G_1 \rightrightarrows G_0$ is a family of positive measures $\{\mu_{G}^{x}\}_{x \in G_0}$ such that $\mu_G^{x}$ is a Lebesgue measure on $t^{-1}(x)$ satisfying the following properties:
	\begin{enumerate}
		\item The family is invariant under left translations,
		\item For all compactly supported function $f \in C^{\infty}(G_1)$ the map
		\beq
			x \mapsto \int_{t^{-1}(x)} \mu^x_G(g) f(g)
		\eeq
		is smooth. 
	\end{enumerate}
	In \cite{landsman1999lie,paterson2012groupoids} it is shown that every Lie groupoid possesses a left Haar system.  
\end{definition}

Given a Lie groupoid $G = G_1 \rightrightarrows G_0$, we can now associate to it a canonical $C^*$ algebra called the groupoid algebra, denoted by $C^*(G)$.\footnote{Strictly speaking, this presentation of $C^*(G)$ is not canonical since it depends on the choice of a left invariant Haar system. There is a more formal construction which avoids this -- see e.g. \cite{landsman2000quantization}. Nevertheless, the more formal construction is much less transparent, so we have opted for the current presentation.} 

\begin{definition}[Groupoid algebra]
	Let $G = G_1 \rightrightarrows G_0$ be a Lie groupoid with left Haar system $\{\mu_{G}^x\}_{x \in G_0}$. To begin, we will consider the set of compactly supported functions on $G_1$, $C^{\infty}_c(G_1)$, and seek to endow on this set the structure of an involutive algebra. For the product structure on $C^{\infty}_c(G_1)$ we use the convolution of maps:
	\beq
		\bigg(a \star b\bigg)(g) \equiv \int_{t^{-1}(s(g))} \mu_G^{s(g)}(g') \; a(gg')b(g'^{-1}), \; a,b \in C^{\infty}_c(G_1).
	\eeq
	For the involution, we use
	\beq
		(a^*)(g) = \overline{a(g^{-1})}.
	\eeq
	Finally, to turn $C^{\infty}_c(G_1)$ from a $*$-algebra to a $C^*$-algebra we must select an appropriate $C^*$ norm and complete $C^{\infty}_c(G_1)$ in its associated topology. A standard choice is
\beq
	\norm{a} \equiv \text{max}\bigg\{\sup_{x \in G_0} \int_{t^{-1}(x)} \mu^{x}_{G}(g) |a(g)|, \sup_{x \in G_0} \int_{t^{-1}(x)} \mu^{x}_{G}(g) |a(g^{-1})| \bigg\}.
\eeq	 
The resulting algebra is what we call $C^*(G)$. 
\end{definition}

Given a Lie groupoid with Haar system $\{\mu^{x}_G\}_{x \in G_0}$, it is mechanically straightforward to generalize the definitions given in Section \ref{sec: CP}. The group automorphism must be promoted to a groupoid automorphism, which can be encoded in terms of a groupoid representation, and the space $L^2(G;\mu)$ should be interpreted as the set of square integrable functions with respect to the Haar system. For a more detailed discussion of crossed products of $C^*$ algebras by locally compact topological groupoids, we refer the reader to \cite{khoshkam2004crossed,anantharamandelaroche2021remarks}.

\bibliographystyle{uiuchept}
\bibliography{CPI_JournalRevision}
\end{document}